\documentclass{biophys-new}

\usepackage[utf8]{inputenc}
\usepackage{graphicx}

\usepackage{lipsum}

\usepackage{soul}

\usepackage{todonotes}

\usepackage{siunitx}

\usepackage[colorlinks,allcolors=cyan!70!black]{hyperref}

\title{Deformability and collision-induced reorientation enhance cell topotaxis in dense microenvironments}
\runningtitle{Cellular properties enhance topotaxis} %% For page header

\author[1,7,*]{Leonie van Steijn}
\affil[1]{Mathematical Institute, Leiden University, Leiden, The Netherlands}
\author[2]{Joeri A. J. Wondergem}
\affil[2]{Leiden Institute of Physics, Leiden University, Leiden, The Netherlands}
\author[1,2]{Koen Schakenraad}
\author[3,4,5]{Doris Heinrich}
\affil[3]{Fraunhofer Institute for Silicate Research ISC, 97082 W\"urzburg, Germany}
\affil[4]{Institute for Bioprocessing and Analytical Measurement Techniques, 37308 Heilbad Heiligenstadt, Germany}
\affil[5]{Faculty for Mathematics and Natural Sciences, Technische Universität Ilmenau, 98693 Ilmenau, Germany}
\author[1,6]{Roeland M. H.  Merks}
\affil[6]{Institute of Biology, Leiden University, Leiden, The Netherlands}

\runningauthor{Van Steijn et al.} %% For page header
\affil[7]{Present address: Theoretical Biology \& Bioinformatics, Universiteit Utrecht, Utrecht, The Netherlands}
\corrauthor[*]{l.vansteijn@uu.nl}

\papertype{Article}
\makeatletter
\newcommand*{\addFileDependency}[1]{% argument=file name and extension
  \typeout{(#1)}
  \@addtofilelist{#1}
  \IfFileExists{#1}{}{\typeout{No file #1.}}
}

\def\ttl@useclass#1#2{%
  \@ifstar
    {\ttl@labelfalse\@dblarg{#1{#2}}}% {\ttl@labelfalse#1{#2}[]}%
    {\ttl@labeltrue\@dblarg{#1{#2}}}}
\makeatother

\usepackage{setspace}

\begin{document}
% \myexternaldocument{SupplementaryFigures}
% \doublespacing
\begin{frontmatter}

\begin{abstract}
\textit{In vivo}, cells navigate through complex environments filled with obstacles like other cells and the extracellular matrix. Recently, the term 'topotaxis' has been introduced for navigation along topographic cues such as obstacle density gradients. Experimental and mathematical efforts have analyzed topotaxis of single cells in pillared grids with pillar density gradients. A previous model based on active Brownian particles (ABP) has shown that ABPs perform topotaxis, i.e., drift towards lower pillar densities, due to decreased effective persistence lengths at high pillars densities. The ABP model predicted topotactic drifts of up to 1\% of the instantaneous speed, whereas drifts of up to 5\% have been observed experimentally. We hypothesized that the discrepancy between the ABP and the experimental observations could be in 1) cell deformability, and 2) more complex cell-pillar interactions.
Here, we introduce a more detailed model of topotaxis, based on the Cellular Potts model.
To model persistent cells we use the Act model, which mimicks actin-polymerization driven motility, and a hybrid CPM-ABP model. Model parameters were fitted to simulate the experimentally found motion of \textit{D. discoideum} on a flat surface. For starved \textit{D. discoideum}, the topotactic drifts predicted by both CPM variants are closer to the experimental results than the previous ABP model, due to a larger decrease in persistence length. Furthermore, the Act model outperformed the hybrid model in terms of topotactic efficiency, as it shows a larger reduction in effective persistence time in dense pillar grids. Also pillar adhesion can slow down cells and decrease topotaxis. For slow and less persistent vegetative \textit{D. discoideum} cells, both CPMs predicted a similar small topotactic drift. We conclude that deformable cell volume results in higher topotactic drift compared to ABPs, and that feedback of cell-pillar collisions on cell persistence increases drift only in highly persistent cells.
\end{abstract}

\begin{sigstatement}% word limit 120 words
Knowing how the environment influences cell motility is useful in developing methods to interfere during disease or in tissue engineering. One factor is the presence of obstacles: in a process called topotaxis single cells move from a high to a low density of obstacles. Here, we show that a number of cellular properties, namely deformable volume, contact inhibition of locomotion, and adhesiveness to obstacles, influence the efficiency of topotaxis. Understanding the differences in these properties between cell types could point to cell-sorting mechanisms for tissue-engineering, or shed light on the migratory behavior of immune and cancer cells.
\end{sigstatement}
\end{frontmatter}

\section*{Introduction}
Motile cells moving through tissue encounter many obstacles, such as other cells and the extracellular matrix. How cells react to the density of obstacles is important for many biological processes, such as cancer cells invading from a dense tumor into looser packed tissues \cite{Pathak2012,Ranamukhaarachchi2019MacromolecularMorphogenesis}, immune cells moving through tissues with different porosity \cite{Krummel2014ModesMyosin-IIA,Toyjanova2015} or pathogens such as \textit{Plasmodium} that migrate through different tissues throughout their life cycle \cite{Hellmann2011,Battista2014}. Recently, various \textit{in vitro} assays provided experimental evidence that cells use the topography of their environment to actively orient themselves \cite{Park2016,Heo2017,Mahmud2009,Caballero2015,Hellmann2011,Reversat2020CellularTopography}. Although topographically induced cell migration has predominantly been shown for \textit{in vitro} environments, it is likely to also play a role \textit{in vivo}, as the typical cellular environment contains a rich variety of physical structures ranging from the nano- to milimeter scale.
% in development, where cells migrate through developing tissues to reach their destination site. But also after development topotaxis can play a role in the immune system, where immune cells move through the tissues to clear invaders, and in cancer invasion and metastasis.
% \todo[inline, author=Leonie]{Elaborate. Plays role in immune system, cancer invasion}

\textit{Topotaxis} is an excellent umbrella term to describe all such forms of biased cell migration caused by asymmetries in the topography of the cellular environment. Recently, the term was reintroduced by Park et al. to describe topographic cell guidance by sub-cellular pillar geometries \cite{Park2016}. Such assays, with topotactic cues at the subcellular scale, result in uninterrupted physical feedback between the cell body and the topographic asymmetry, which was shown to result in biased motion for various configurations \cite{Park2016,Heo2017,Mahmud2009,Caballero2015}.
% Mahmud Key players in this bias were the rearrangement of actin and formation of the lamellipodium.
However, for highly motile persistent-random cell motion, topotaxis can also be generated by topotactic cues which span distances much larger than the typical cell body, as shown by Wondergem et al. \cite{Wondergem2019}. In these experiments, \textit{Dictyostelium discoideum}, a model organism used to study persistent random cell motion \cite{Li2011}, was shown to drift into areas of lower obstacle density and increased spacing over long distances. The spacing between subsequent rows of obstacles (pillars of cell size) was only increased after multiple cell lengths, introducing changes in topography at distances beyond the cell size. Hence, in these experiments the cell body was not in continuous contact with the topographic asymmetry, in contrast to the nanoscale and microscale structures described in \cite{Park2016,Caballero2015}. Yet, the asymmetric pillar grid provided a sufficiently strong topographic cue to cause a net drift of approximately 5\% of cell speed over large distances, even when exposed to competing chemotactic cues.

For such long distance topotaxis to occur, Schakenraad et al. \cite{Schakenraad2020} hypothesized that the drift could be the result of local differences in the effective persistence of cell motion, which is continuously modified by various cell-obstacle interactions. Using active Brownian particles (ABPs) as a model system for cell motion, they showed this is indeed the case. Particle trajectory simulations in anisotropic pillar fields resulted in local differences in persistence and to a net topotactic drift \cite{Schakenraad2020}. Such positionally dependent kinesis was first suggested as a contributing cause for bacterial chemotaxis\cite{Berg1972ChemotaxisTracking}, where cells tumble more frequently in lower chemoattractant concentrations, and for observed durotaxis on stiffness gradients \cite{Novikova2017}, where cells on softer substrates change direction more slowly than on stiffer substrates. However, the ABP model underestimates the topotactic effect measured for live-cells, finding a 1\% drift when compared to particle speed, which is lower than the 5\% observed for \textit{Dictyostelium}. The ABP model simulated the cell as an undeformable disk moving at a constant speed and mediated particle-pillar interactions through a simple hard wall force. Both choices allow for a very intuitive study into the role of persistence in topotactic drift, but leave out any intricacies posed by the dynamically changing cell body while navigating complex environments.

Here, we explore more refined models of cell migration, and, more specifically, investigate if either the deformable cell body or more complicated cell-pillar interactions play a pivotal role in long-distance topotaxis. The cell body likely influences migration through the obstacle fields, as the cell needs to undergo various morphological changes to squeeze through the pores between pillars. Moreover, when moving through the pillar field, cells are not guided around pillars at constant velocity by a hard-wall force as ABPs, but, for example, reorient more similar to Contact Inhibition of Locomotion (CIL). In order to investigate different types of cell-pillar interactions, and gauge the influence of the cell body, we tested two different methods of persistent-random motion within the framework of the Cellular Potts model (CPM). The first method is based on the Act model, which models cytoskeletal rearrangements \cite{Niculescu2015}. This actin-extendend CPM allows for emergent, active reorientation, as the cell cannot extend into a pillar and will be nudged to repolarize after the interaction, similar to CIL. The second method is a hybrid ABP-CPM model, which excludes any possibility of CIL, as the direction of preferred motion of the simulated cell changes solely based on an inherent characteristic persistence time, but irrespective of obstacle interactions, similar to the ABP model. Although pillars are unable to alter the inherent persistence in the ABP-CPM, the simulation does retain any influence of the deformable cell body on cell migration through the pillar grid. The parameters of the CPM, and both associated methods of driving persistent random motion, were empirically calibrated by measuring \textit{Dictyostelium discoideum} migration on flat surfaces.

The CPM simulations of the two methods of persistence resulted in a different topotactic response. The ABP-CPM simulations resulted in more topotactic drift than ABPs for the same motion parameters, clearly demonstrating the influence of a deformable cell body on the efficiency of topotactic response. The Act-CPM simulations exhibited the most pronounced topotactic drift, which we show can be attributed to changes in both speed and persistence by simulating cell motion in pillar grids of constant spacing. To compare the simulations to experimental data, we measured the topotactic drift for starved \textit{D. discoideum} migrating on the pillar field introduced by Schakenraad et al. \cite{Schakenraad2020}, which allows for a continuously changing linear gradient, contrary to earlier experimental work. Measuring cell migration on the same lattices used for modeling topotaxis ensured a more optimal comparison between experimental and simulated data. The ABP-CPM most closely matches the experimental topotactic drift, whereas the Act-CPM slightly overpredicts topotaxis. However, when normalizing for cell speed, the measured topotactic drift of \textit{D. discoideum} sits in-between the ABP-CPM and Act-CPM simulated drifts. Finally, we tested the effect of changing the adhesion energy between cells and pillars. Changing this adhesion energy has a direct influence on the cell speed next to pillars, and thereby topotactic drift, and equates to varying the surface material of the pillars in \textit{in vitro} assays.

\section*{Materials and Methods}\label{sec:M&M}

For this study, we have used an experimental setup with live cells and three computational models. Experimental data from cells on a flat surface were used to set the model parameters and calibrate the simulations. Topotaxis caused by gradients of anisotropically placed cell-sized obstacles (pillars) was investigated in experimental assays, as well as in computational models.

% \subsection*{Pillar gradient}

\subsection*{Live cell experiments}

	\subsubsection*{Cell culture and experiment preparation}\label{sec:CC_and_LiveExp}
	For all migration experiments, Axenic \textit{D. discoideum} (Ax2) with a cytoplasmic green fluorescent protein (GFP) insertion was used (strain HG1694, obtained from Dr. G. Gerisch, MPI for Biochemistry, Germany). Cells were grown at  \SI{20}{\celsius} in HL5 medium, supplemented with \SI{20}{\micro g.ml^{-1}} gentamicin (Gentamycin solution, Merck, Netherlands) as a selection antibiotic. The cells were cultured in 100 mm petri dishes (100 mm TC-treated culture dish, Corning, USA) and confluency was kept below 70\% during culturing. For visualizing  actin  polymerization,  a  LimE-GFP  in  lim0  cell line  was  used \cite{Pang1998}.   These  cells  were  cultured similarly, but with two selection antibiotics instead, 10 \si{\micro g.ml^{-1}} Gentamycin and 10 \si{\micro g.ml^{-1}} Blasticidin, necessary to maintain the double mutation (LimE-GFP and Lim0).

	In preparation for imaging experiments, cells were harvested by pipet induced flow and collected in a conical tube. To remove the culture medium, the cells were centrifuged at 1500 rpm for 3 min. In case of vegetative experiments, resulting cell pellets were thrice washed using non-fluorescent buffer (3.6mM KH$_{2}$PO$_{4}$, 2.9mM Na$_{2}$HPO$_{4}$, PH 6.7) and, after resuspension, transferred onto (un)structured polydimethylsiloxane (PDMS) surfaces placed inside an imaging chamber (see Sec. \nameref{sec:MM_obs_surface}). In case of starved experiments a pulsation procedure was started before imaging instead. Cell pellets were thrice washed with 17 mM K-Na-phosphate buffered saline (PBS, pH 6.0) and placed on a shaker for 1 hour. Then, to induce cAR1 expression, cells were pulsed with 150 nM cyclic adenosine-monophosphate (cAMP, Merck, The Netherlands) applied in 6 minute intervals over 4 hours while shaking. After pulsation, any residual cAMP was removed by centrifugation and resuspension. Cells were left to shake in a conical tube with PBS for another 30 min before being loaded onto PDMS surfaces in imaging chambers. For both experiments cells were left to adhere for $1$ hour after insertion into the imaging chamber, leading to a 6-7 hour starvation period. Cell seeding concentrations were kept below $1 \cdot 10^{-4} $ \si{\micro m^{-2}} to limit cell-cell interaction and enable studying migration of individual cells.

	\subsubsection*{Obstacle and flat PDMS surfaces}
	\label{sec:MM_obs_surface}
	PDMS (Sylgard 184 Silicon Elastomer Kit, Dow Corning, USA) was mixed 1:10 resulting in a 1.72 MPa stiffness for all surfaces \cite{Palchesko2012}. Flat surfaces were prepared by spin-coating pre-mixed PDMS onto polished Silicon wafers (Siegert Wafer, Germany) and then baked for 4 hrs at 110 \si{\degree}C. Before spin-coating, wafers were silanized by Trichloro(1H,1H,2H,2H-perfluorooctyl) silane deposition under vacuum (50 mbar) for 1 hour to ensure proper PDMS detachment later. Before use, PDMS was cut, peeled off, and washed with ethanol (70\%).  For  migration experiments  PDMS  was  cast  around a 250-500 \si{\micro m} thickness, for limE-GFP imaging, PDMS was cast ultra-thin ($<$50\si{\micro m}) enabling 100x (WD= 130\si{\micro m}) imaging.

	Pillar obstacle fields were prepared using a molding process. The pillar molds were prepared by two-photon direct laser writing (DLW) using the Photonic Professional GT (Nanoscribe, Germany). First, a negative of the topotaxis pillar grids ($s=0.01\mbox{-}0.03$, see Subsection \nameref{def:pillars}) were designed using Inventor (Autodesk, USA) and, via a stereolithography format (.stl), imported to DeScribe (Nanoscribe, Germany) to prepare for DLW. Then, two-photon crosslinking was performed using the IP-S resin (Nanoscribe, Germany) deposited on a silicon wafer. Different laser powers and scan speeds were chosen for bulk and edges of the structure, 27\% and 42\% (of 140 mW), 10 and 40 \si{\milli m/min} respectively. To remove excess resin, molds were developed for 45 min in polyglycidylmethacrylate (PGMA). After blow drying with nitrogen, wafers were silanized as described previously. To produce the pillar field casts, PDMS was deposited over the wafer (with mold on top), baked, cut, peeled off, and washed with ethanol (70\%).

	PDMS inserts were hydrophilicitized by 15 min of UV/Ozone exposure (UVO-42, Jelight Company, U.S.A) and placed inside an imaging slide (0.8 sticky-Slide I Luer, Ibidi, Germany), to be used immediately. Before loading cell suspensions, imaging slides were washed with ethanol and then PBS. To determine quality of the mold and PDMS inserts, a cast of each pillar design was imaged using a nanoSEM (FEI/Thermo Fisher, The Netherlands) scanning electron microscope (SEM). Samples were imaged at 10kV, with a spot size of 4.0. Before imaging, PDMS structures were coated with 2-8 nm Pt/Pa using a plasma magnetron sputter coater (208HR, Cressington, Watford, UK) to enhance conductance.

	\subsubsection*{Live-cell imaging and tracking}
	\label{sec:Imaging_tracking}
	Cells were imaged every 8-10 seconds for experiments on flat PDMS and every 20 seconds for topotaxis assays. Measurements lasted for 1-3 hours and were performed with either 10x or 20x air objectives (Plan Fluor, Nikon) on a Nikon Eclipse Ti microscope equipped with a confocal spinning disk unit operated at 10,000 rpm (Yokogawa). The cytoplasmic GFP was excited at 488nm by a solid-state diode laser (Coherent, U.S.A.) supported in an Agilent MLC4 unit (Agilent Technologies, U.S.A.), at reduced intensity (25\% of 2.4mW) controlled by an Acousto-Optic Tunable Filter. Emission was filtered by a quad-band fluorescence filter (TR-F440-521-607-700, IDEX LLC, Rochester, New York, U.S.A.). Images were captured using an exposure time of 200ms by an Andor iXon Ultra 897 High Speed EM-CCD camera (Andor Technology, U.K.). Images of higher  magnification were produced with the same setup but using different objectives: Figure \ref{fig:IVTopotaxis}E with 40x air (PlanFluor, Nikon), limE-GFP Figures \ref{fig:IVTopotaxis}D and \ref{fig:collision}A with 60x water (Plan Apo VC, Nikon) and Figures  \ref{fig:TuningCPM}C and \ref{fig:TuningCPMII}A with 100x oil (CFI plan Apo, Nikon).

	Image tracking was performed using ImageJ (http://imagej.nih.gov/ij/). Microscopy time-lapse images were contrast and brightness adjusted, and ran through a Gaussian filter ($\sigma = 2$) to enable optimal cell body recognition. The ImageJ plugin CellEvaluator was used to determine the $x,y$-coordinates of the center-of-mass of each cell body in each frame \cite{Youssef2011}. By linking all these $x,y$-positions together, cell trajectories were obtained.

	\subsubsection*{Empirical measurement of cell area}
	Fluorescence microscopy images of cell migration experiments on flat PDMS were used to fit the cell surface area ($A_{\sigma}$) parameter in the CPM. Frames used for area analysis were chosen sufficiently far apart ($f^{-1}=\delta t = $ 200 s) to avoid correlations. The two-dimensional projection of the cell body was determined using Sobel edge detection, applied to cytoplasmic GFP images obtained using the 20x air objective (0.657 \si{\micro m/pix}). Brightness and contrast were adjusted to rescale pixel intensities (i.e. only use the range of 16-Bit intensities of GFP signal detected). Image analysis was performed using Matlab, and the Image Processing Toolbox (Matlab v2019a, MathWorks, U.S.A.) in particular. First, edges were detected (Sobel) using an appropriate threshold, then the resulting binary edge-image was dilated, holes were filled, borders cleared and the image eroded (equal to the initial dilation). Then all groups of non-zero, adjacent pixels forming a cell were identified and properties (using \textit{regionprops}), like the area, extracted.

\subsection*{Models}\label{sec:Mod}
\subsubsection*{ABP model}
We use the ABP model as previously described by Schakenraad et al. \cite{Schakenraad2020}. In this model, a particle is described as a disk of radius $R_p$ at location $\mathbf{r}(t)$, with speed $v_0$ moving in the direction $\mathbf{p}=(\cos \theta, \sin \theta)$ which change according to:
\begin{eqnarray}
    \frac{d\mathbf{r}}{dt} = & v_0\mathbf{p} +\mu\mathbf{F}\\
    \frac{d\theta}{dt} = & \sqrt{2D_r}\xi, \label{eq:ABP-ode}
\end{eqnarray}
with $\mu$ a mobility coefficient, $\mathbf{F}=\mathbf{F(r)}$ the interaction force between particle and obstacle, $\xi=\xi(t)$ a random variable with zero mean and time correlation $\langle\xi(t)\xi(t')\rangle=\delta(t-t')$.  The rotational diffusion coefficient $D_r=1/\tau_p$ is the inverse of the persistence time $\tau_p$ of the ABP. The interaction force $\mathbf{F}$ is given by
\begin{equation}
\mathbf{F}=
    \begin{cases}
    -\frac{v_0}{\mu}(\mathbf{p}\cdot\mathbf{N})\mathbf{N},& \text{if } |\Delta\mathbf{r}_0|\leq R\\
    0,              & \text{otherwise},
\end{cases}
\end{equation}
with $\mathbf{N}$ a unit vector normal to the obstacle surface, $|\Delta\mathbf{r}_0|$ the distance between the centers of the particle and the obstacle, and $R$ the effective particle radius as the sum of the particle and pillar radius $R=R_p+R_0$. In our simulations $\mu=1$, and $v_0$ and $\tau_p$ are set to experimentally determined cell speeds and persistence times. $R_p=$\SI{1}{\micro\metre} by default, to represent the size of the nucleus of $\textit{D. discoideum}$ \cite{Stevense2011NuclearDictyostelium,Belotti2021EffectsCells}, or is set to \SI{0}{\micro \metre} (no volume), or \SI{1.8}{\micro\metre}, which is the maximal possible radius to fit through the smallest distance in the topotaxis grid.
% \todo[inline]{Hier eerst een overzicht geven van de opzet van de hele modelstudie, in niet-(al-te)-technische termen. Dus een model van flexibele persistent-randomly walking cells, die vervolgens worden 'losgelaten' in een arena met gesimuleerde pilaren. Voeg ook een figuur toe van de modelopzet met pilaren en een beweeglijke cel. Daarna kort beschrijven hoe je de cellen gaat analyseren}

\subsubsection*{Cellular Potts model of persistently moving cells}
We use the 2D Cellular Potts model (CPM) \cite{Graner1992SimulationModel} with either of two extensions for modeling persistent cell movement (see Paragraphs \nameref{par:act} and \nameref{par:abpcpm}). The CPM represent cells on a regular square lattice $\Lambda \subset \mathbb{Z}^2$. Each lattice site, $\vec{q}\in\Lambda$, is associated with a spin value $\sigma( \vec{q}) \in \{1,0,-2\}$, that assigns  the lattice site to the cell ($\sigma( \vec{q})=1$), the medium ($\sigma( \vec{q})=0$), or a pillar ($\sigma( \vec{q})=-2$).
% \todo{etc. etc. the modelbeschrijving is nog erg 'losjes' geschreven. Laat je inspireren door hoe dit gaat in vorige artikelen. Hier en hieronder bij "pillars" wat eerste suggesties om dit exacter op te schrijven}
The cell is represented as the collection of lattice sites marked with $\sigma=1$: i.e., $\{\vec{q}\in\Lambda|\sigma(\vec{q})=1\}$.
% Lattice sites that are not occupied by a cell represent medium or a pillar and are indicated with $\sigma(medium)$ or $\sigma(pillar)$.

Cell motion is modeled by updating the lattice through iterative attempts to extend or retract the cell into neighboring lattice sites. For a copy attempt, the algorithm  selects at random a lattice site $\vec{q}$ and one of its eight neighbors $\vec{q'}$. If $\sigma(\vec{q})\neq\sigma(\vec{q'})$, the algorithm computes $\Delta\mathcal{H}_\text{total}$, the change in Hamiltonian together with the energy change due to active forces from the cell's persistence, which determines the success of a copy attempt. The Hamiltonian $\mathcal{H}$ balances the forces of cell-medium and cell-pillar interactions, and cell area and cell perimeter constraints \cite{Ouchi2003ImprovingCells}:
% \begin{equation}
\begin{eqnarray}
\mathcal{H}& = & \sum_{\vec{u},\vec{v}} J_{\sigma(\vec{u}),\sigma(\vec{v})}(1-\delta_{\sigma(\vec{u}),\sigma(\vec{v})}) + \lambda_{\text{area}}(a_\sigma-A_\sigma)^2 \nonumber \\
&&+ \lambda_{\text{perimeter}}(p_\sigma-P_\sigma)^2
\end{eqnarray}
% \end{equation}
Here, $J_{\sigma(\vec{u}),\sigma(\vec{v})}$ describes the adhesion energy between two neighboring lattice sites $\vec{u},\vec{v}$ of types $\sigma(\vec{u})\neq\sigma(\vec{v})$. We set $J_{\text{pillar,medium}}=0$ to only take the adhesion energy between cell and medium ($J_{\text{cell,medium}}=20$ \cite{Niculescu2015}) or cell and pillar ($J_{cell,pillar}$, varies) into account. The second and third Hamiltonian term describes the area  and perimeter constraint and penalize deviation of the cell area $a_\sigma$ from its target area $A_\sigma$ and of cell perimeter $p_\sigma$ from the target perimeter $P_\sigma$. The $\lambda$'s indicate the weight of both constraints.

The probability of a successful copy attempt depends on $\Delta\mathcal{H}_\text{total}=\Delta\mathcal{H}-\Delta\mathcal{H}_\text{motility}$:
 $$					     P(\Delta\mathcal{H}_\text{total}) =
					\left\{
						\begin{array}{ll}
							1  & \mbox{if } \Delta\mathcal{H}_\text{total} < 0 \\
							e^{-(\Delta\mathcal{H}_\text{total}/T)} & \mbox{if } \Delta\mathcal{H}_\text{total} \geq 0
						\end{array}\right.,
$$

with $T$ denoting the noise in the system that allows for energetically unfavorable copy attempts to be accepted. Model time is expressed in Monte Carlo steps (MCS). Within a single MCS, the expectation is that for each lattice site an update has been attempted once. Since we model only a single cell in a large field, many neighboring lattice site pairs will not lead to cell movement. We ignoring such unfruitful copy attempts by using a rejection-free algorithm to speed up simulations \cite{Lee1995Rejection-freeTechnique,Starru2014}.

% single Monte Carlo step is a number of copy attempts. We define it using the set of lattice sites pair that would lead to actual changes in the lattice coloring. Hence, we take the set edges, or neighboring lattice sites $x$ and $x'$, such that $\sigma(x)\neq\sigma(x')$, excluding edges between medium and pillars, and divide that number by the neighborhood size. Our definition of a Monte Carlo step is also dynamic: if a cell grows during a MCS, the number of edges will also increase, which will take into account by extending the number of copy attempts for that Monte Carlo step, and similarly for shrinking cells.\todo{Gebruik je hier het edgelist algoritme door Martijn?}.

\paragraph{Act-CPM}\label{par:act}
In order to model persistent cells, we use two different extensions of the CPM. First, we use the Act-CPM \cite{Niculescu2015}. It models the actin polymerization driven persistent cell motility \cite{Niculescu2015}. Each lattice site $\vec{q}$ has an extra value, $Act(\vec{q})$, which can range from 0 to $\text{Max}_\text{Act}$ inside the cell . The $Act$-value outside the cell is always 0. If the cell has recently made an extension into $\vec{q}$, $Act(\vec{q})$ is set to $\text{Max}_\text{Act}$. Every Monte-Carlo step, each non-zero $Act$-value will be decreased by 1, until 0.

For the Act-CPM, $\Delta\mathcal{H}_\text{motility}=\Delta\mathcal{H}_{Act}$. In a copy attempt extending the cell from a lattice site $\vec{c}$ into an empty lattice site $\vec{m}$, we look at the geometric mean of the $Act$-values in $\text{NB}(\vec{c})$, the Moore neighbors of $\vec{c}$ that are also within the cell, such that:
\begin{equation}\Delta\mathcal{H}_{Act}(\vec{c}\rightarrow \vec{m})=\frac{\lambda_{Act}}{\text{Max}_{Act}}  \left(\prod_{\vec{y}\in \text{NB}(\vec{c})}Act(\vec{y})\right)^{\frac{1}{|\text{NB}(\vec{c})|}}.
\end{equation}  For retraction attempts, the sign changes, so $\Delta\mathcal{H}_{Act}(\vec{m}\rightarrow \vec{c})=- \Delta\mathcal{H}_{Act}(\vec{c}\rightarrow \vec{m})$. As a consequence,  the cell is more likely to extend outwards at areas with high $Act$ values and is less likely to retract from such areas. This simulates the polarized actin structure of a cell in a phenomenological way. The two parameters $\lambda_\text{Act}$ and $\text{Max}_\text{Act}$ can be tuned to obtain different cell motilities.

\paragraph{ABP-CPM}\label{par:abpcpm}
The ABP-CPM is a hybrid method between a CPM model extension \cite{Beltman2007} and the ABP model \cite{Schakenraad2020}. Like the ABP model, the cell has a direction $\mathbf{p}=(\cos\theta,\sin\theta)$. Cell movement along this direction is favored. For each copy attempt extending the cell, the angle $\alpha$ between the resulting displacement of the center of mass and $\mathbf{p}$ is computed, and results in $\Delta\mathcal{H}_\text{motility}=\Delta\mathcal{H}_\text{persistence}$ like in \cite{Beltman2007}:
\begin{equation}
    \Delta\mathcal{H}_{\text{persistence}}=\lambda_{\text{persistence}}\cdot\cos(\alpha).
\end{equation}

For updating $\theta$, we use the same differential equation as in \cite{Schakenraad2020}, see Eq.~\ref{eq:ABP-ode}:
\begin{equation}
    \frac{d\theta}{dt}=\sqrt{\frac{2}{\tau_p}}\xi(t).
    \label{eq:ODE}
\end{equation}
Here $\tau_p$ stands for the persistence time, and $\xi(t)$ is a stochastic white noise term, modeled by a Gaussian distribution with mean 0 and variance $\sigma^2=2\frac{\Delta t}{\tau_p}$.
The term $\Delta t$ couples the time of equation \ref{eq:ODE} to the MCS, and we choose $\Delta t=0.1\tau_p$. At initialization, the cell is assigned a $\theta$ from a uniform random distribution. We update $\theta$ each MCS. Different cell motilities can be obtained by tuning the parameters $\tau_p$ and $\lambda_{\text{persistence}}$.
% \todo[inline, author=Leonie]{Find out what $\Delta t $ actually does. Do we need it to be variable or can it be set? Does is alter time perception?}
% \todo[inline,author=Leonie]{Implement}
% \todo[inline,author=Leonie]{Write about implementation}

\subsubsection*{Pillars}
\label{def:pillars}

%\subsubsection*{Pillar grid with gradient}
We utilize two different pillar grids: a regularly spaced pillar grid and a pillar grid with a topotaxis gradient. The regularly spaced grids are defined by the distance between pillar centers $d$ and the pillar radius $R$. The topotaxis grid is defined as by Schakenraad et al.\cite{Schakenraad2020}. It consist of three different parts: a regularly spaced part on the left of the field, a part with a gradient in pillar distance in the middle and another regularly spaced part on the right.
The pillar gradient in the middle is defined by the following set of pillar centers $P$,

\begin{equation}
  P =\left\lbrace (x,y)\in \mathbb{R}^2 \;\middle|\;
  \begin{tabular}{@{}l@{}r}
   $x=\frac{d}{1-e^{-s}}(e^{sn}-1)+\frac{d}{2}$& and\\
     $y=d\left(m+\frac{1}{2}\right)e^{sn}$,& with $n, m\in \mathbb{Z}$
   \end{tabular}
  \right\rbrace,
  \label{eq:pillarcenters}
\end{equation}
where $d$ is the distance between pillar centers at the origin $(0,0)$, the center of the field.
% \hl{in the origin, the center of the field}\todo{Bedoel je hier at $(0,0)$, i.e, the center of the field?}.
This gives a gradient in the horizontal axis with parameter $s$ defining the steepness. The two regularly regularly spaced grids at the left and right of the gradient part of the grid with respective pillar spacings $d_{\min}$ and $d_{\max}=2d-d_{\min}$ limit the gaps between pillars to get too small or too big. Table~\ref{tab:gridpar} gives an overview of all grid measurements.

% \todo[author=Leonie]{What sizes in experiments? 13,65 and 19,95 \si{\micro\metre}? }
% We have determined the minimal required distance between the outsides of the pillars to be 6 grid points, hence we set $d_{\min}=2R+6$. The maximum distance between pillars is determined by $d_{\max}=2d-d_{\min}$
%\todo[author=Leonie]{Recompute $d_{\min}$, with non-vacuum pillars.}Done, doesn't change
%\todo[author=Leonie]{Explain $d_{\max}$}

% \footnote{Minimal pillar distance is 26 for dicty\_empty\_14 parameter set.}

In the CPM, a lattice site
% \todo{in CPM hebben we het meestal over lattice sites, $\vec{x}$. Nieuwe synoniemen maken het lastiger te volgen}
$\vec{x}$ is assigned to a pillar if it is within a distance $R$ from one of the pillar centers. Pillar lattice sites have $\sigma=-2$, as not to be confused with the $\sigma$s of the medium or cells. As pillars are static, copy attempts into a pillar site are not allowed. However, copy attempts from a pillar site into a cell site $\vec{q}$ are allowed, although, in this case, not $\sigma(\vec{q})=-2$, but $\sigma(\vec{q})=0$ is copied. This allows the cells to retract from the pillars and can be seen as medium flowing back in between cell and pillar from the third dimension which we do not model explicitly. The interaction energy $J_{\text{cell,pillar}}$ can be adjusted to represent adhesive or repulsive pillar surfaces. In this work, $J_{\text{cell,pillar}}=J_{\text{\text{cell,medium}}}$, such that pillar adhesion is neutral in respect to adhesion to the medium, unless stated otherwise.

\subsubsection*{Establishing baseline motility}
In order to quantitatively compare the topotaxis of simulated cells to \textit{D. discoideum} cells, we tried to find parameters such that the simulated cells behave the same as the starved and vegetative \textit{D. discoideum} cells on a flat PDMS surface without any pillars (Figures~\ref{fig:TuningCPM} and \ref{fig:TuningCPMII}). For the ABP model, we set $v_{0}$ and $\tau_p$ directly to the experimentally determined values (see Subsection \nameref{def:trajectories}). For the CPMs, we changed the motility parameters $\lambda_{Act}$ and $\text{Max}_{Act}$, and $\lambda_{\text{persistence}}$ and $\tau_p$ to fit to the experimental data.

The experimental data used for fitting the motility parameters are: the surface area, instantaneous speed, effective diffusion coefficient and persistence time. The surface area and speed were used to determine the length and time scale of the CPMs. The effective diffusion coefficient and persistence time had equal weights in our least squares objective. Using a hill climbing algorithm with multiple restarts, we obtained a shortlist of possible parameter values. The fifteen best parameter sets were then scored ten more times to obtain the best parameter set. The resulting effective diffusion coefficients and persistence times are shown in Tables~\ref{tab:fits},\ref{tab:fitsII}. The optimal parameters are shown in Tables~\ref{tab:par},\ref{tab:parII}, and were used for further simulations.

\subsection*{Trajectory analysis}
\label{def:trajectories}
	Cell trajectories recorded on flat and pillar PDMS surfaces were analyzed with an in-house Matlab code (version 2019b, The Mathworks, U.S.A.). The instantaneous velocities $v$, persistence times $\tau$, and diffusion constants $D$ extracted for the flat trajectory data by this analysis were used to calibrate the CPMs and ABP simulations. Dying or otherwise immotile cells (see \cite{Wondergem2019}) and insufficiently long trajectories (flat $n < 100$, topotaxis $n < 30$) were discarded for analysis, to prevent noise in mean-squared displacement (MSD) values at long-lag times.

	The displacement of the cell between frames was found by $\bar{r}(t) = \bar{R}(t+\Delta T) - \bar{R}(t)$, where $\bar{R}(t)$ are the vectors described by the $x,y$-coordinates of the cell center in each image. The instantaneous velocity of each cell was estimated using the time-averaged velocity between frames,
	\begin{eqnarray}
		\langle v(\Delta T) \rangle_{\textnormal{T}} = \frac{1}{n-1}\sum_{k=1}^{n-1} \frac{\vert \bar{R}(t_k) - \bar{R}(t_k-\Delta T)\vert}{\Delta T}.
	\end{eqnarray}
	Here, the trajectory time $t_k$ is discretized by the constant frame rate of the camera, $t_k = k \cdot\Delta T$, with $k = 0,\dots,n$, $\Delta T = $ \SI{10}{\second} or \SI{20}{\second}, and where $n$ is the number of frames in a trajectory. The ensemble average of each experiment (and simulation) was obtained by taking the mean of all $v$ per experiment. For the experimental trajectories, the time-averaged mean-squared displacement (TAMSD) was used:
	\begin{equation}
	\langle \Delta \bar{r}^{2}(t_{\textnormal{lag}}) \rangle_{\textnormal{T}} = \frac{1}{n-m}\sum_{k=1}^{n-m} \left(R(t_k+t_{\textnormal{lag}}) - R(t_k) \right)^2.
\end{equation}
    In this expression $m$ represents the lag-time multiple under consideration $t_{\textnormal{lag}} = m\Delta T$. The TAMSDs were averaged over all $N$ trajectories per lag-time, and subsequently fit to F\"urths formula,
	\begin{equation}
\textnormal{MSD}(t)=2v_0^2\tau t-2v_0^2\tau^2(1-e^{-t\tau}).
\label{eq:furth}
\end{equation}
	From this fit the characteristic persistence times ($\tau$) and effective diffusion constants ($D_\text{eff}$) for both vegetative and starved \textit{D. discoideum} cells moving on flat PDMS were extracted and used to fit the CPMs and ABP model.

	To determine the persistence time in the simulation, we computed the ensemble-averaged MSD of $N=1000$ simulated cells (CPMs) or $N=10000$ particles (ABP model):
\begin{equation}
    \langle r(t)^2\rangle=\frac{1}{N}\sum_{i=1}^{N}(\bar{R}(t_0+t)-\bar{R}(t_0))^2.
    \label{eq:EAMSD}
\end{equation}
Here, $t_0=0$ for the ABP simulations and $t_0=120 MCS$ for both CPMs to allow for an initialization period for the Act-CPM.
We then fitted to $\langle r(t)^2\rangle$ Fürth's formula (Eq.~\ref{eq:furth}) for the ABP trajectories, or the adapted Fürths formula for the CPM trajectories
\begin{equation}
\textnormal{MSD}(t)=4D_{T}t+2v_0^2\tau t-2v_0^2\tau^2(1-e^{-t\tau}),
\label{eq:furthtransdiff}
\end{equation}
which describes the MSD of a persistent random walk with translational noise \cite{Zeitz2017}.
Here, $\tau$ is the persistence time, $v_0$ the constant speed, and $D_T$ is the translational diffusion caused by thermally induced fluctuation, which is inherent in the CPM. The corresponding effective diffusion constants are described by $D_{\text{eff}}=\frac{1}{2}v_0^2\tau$ (ABP) or $D_{\text{eff}}=D_T+\frac{1}{2}v_0^2\tau$ (CPMs). However, we obtained $D_{\text{eff}}$ independently from Eq.~\ref{eq:furthtransdiff} by fitting the MSD of a Brownian particle
\begin{equation}
\textnormal{MSD}(t)=4Dt,
\label{eq:brownian}
\end{equation}
to Eq.~\ref{eq:EAMSD} for $t\geq2000$ MCS (CPMs) or $t\geq900$ s (ABP):

	For flat and topotaxis assays of the experimental and simulated trajectories, migratory drift ($v_{x,y}$) was calculated by averaging over all displacements of all trajectories,
	\begin{align}
	\langle v_{x,y} \rangle = \frac{1}{n-1} \sum_{i=1}^{n-1} (\bar{R}(t_{i}+\delta t) - \bar{R}(t_{i}))
	\label{eq:drift}
	\end{align}
	where $n$ is the total number of displacements measured. For the simulated topotaxis assays, we performed 1000 single cell CPM simulations or 10000 ABP model simulations for each parameter setting.
	To measure topotactic drift $v_x$ we computed the mean velocity in the $x$-direction with Eq.~\ref{eq:drift}, using the equivalent of $\delta t=20$s intervals for all cell or particles tracks, and $n$ all time steps in which the cell or particle was in the gradient part of the pillar grid.

\section*{Results}

\subsection*{Starved \textit{D. discoideum} cells show topotactic drift increasing with gradient steepness}

\begin{figure*}
		\centering
		\includegraphics[width=\textwidth]{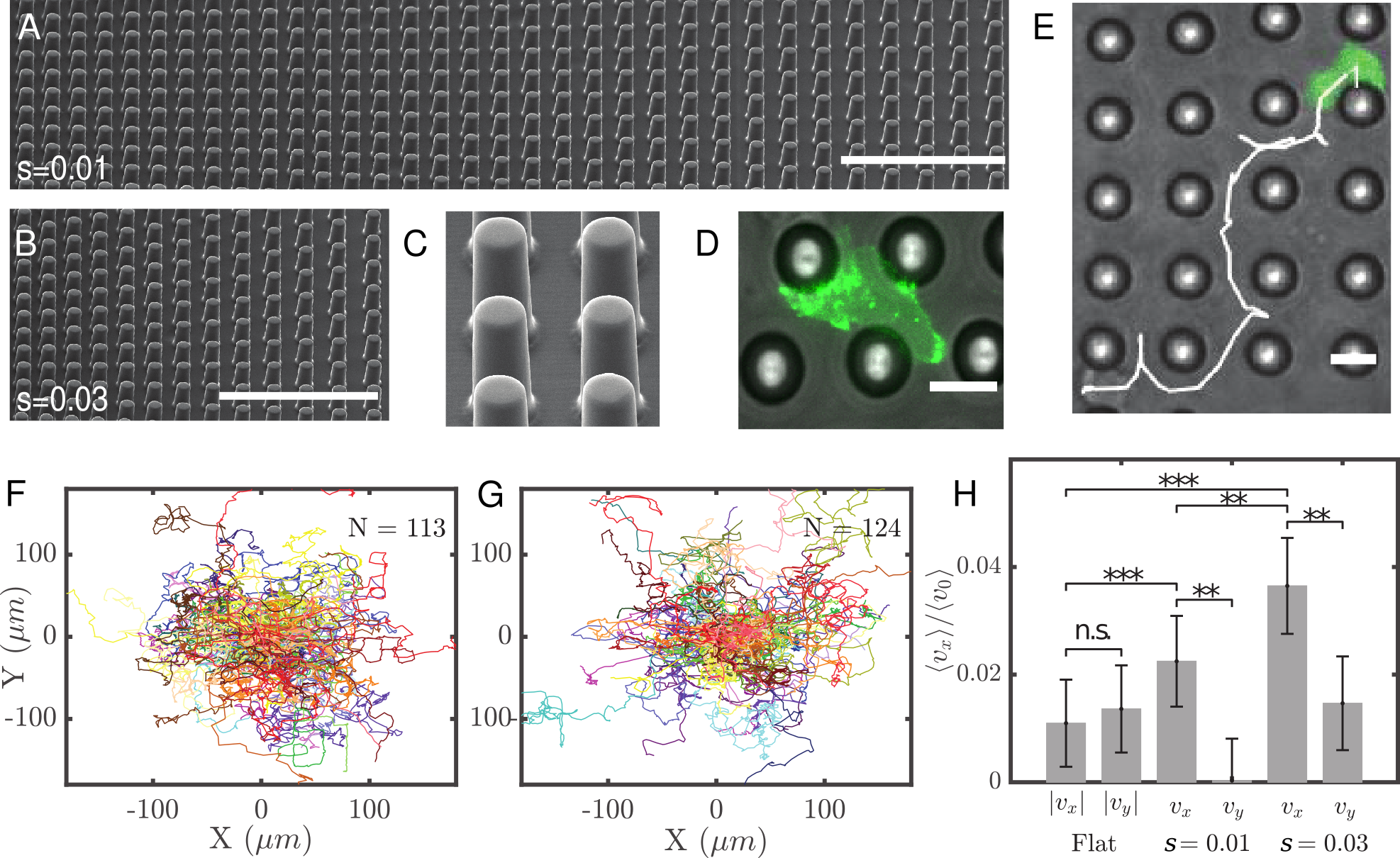}
		\caption{\textbf{Large scale topotaxis for highly-motile amoeboid cell migration (\emph{D. discoideum})} (A) SEM micrograph of the $s = 0.01$ pillar field. Scale bar is 100 \si{\micro\metre}.
% 		Inset shows detailed image of pillars ($h = 20 $ \si{\micro\metre}, $d = 10 $ \si{\micro\metre}). Scale bar is 10 \si{\micro\metre}.
		(B) Same micrograph, but for $s = 0.03$.
% 		SEM micrograph of the $r = 0.01$ pillar field. Scale bar is 100 \si{\micro\metre}.
		(C) Detailed micrograph of pillars ($h =$ \SI{20}{\micro\metre}, $d =$ \SI{10}{\micro\metre}).
		(D) Actin polymerization hot spots visualized for a cell migrating through a pillar field. Fluorescence is LimE-GFP expressed in LimE null cells \cite{Pang1998,Bretschneider2004} after z-projection, scale bar is 10 \si{\micro\metre}. See also \nameref{mov:actvispil}.
		(E) Bright field image of a pillar field ($s = 0.03$) overlaid with the trajectory of a migrating starved \textit{D. discoideum} cell (free cytoplasmic
GFP in green). Scale bar is 10 \si{\micro\metre}.
        (F) Trajectory plot of cells moving on topotaxis field $s = 0.01$.
        (G) Trajectory plot of cells moving on topotaxis field $s = 0.03$.
        (H) Drifts ($v_{x,y}$) measured in all live cell experiments (flat, $s = 0.01$ and $s = 0.03$) compared. The anisotropicity in pillar positioning was in
the $x$ direction for all non-flat experiments. The topotactic drift ($v_x$) was found to be significantly higher ($p \leq 0.01$) than noise ($v_y$) for all topotaxis assays. Additionally, the topotactic drift was found to increase ($p \leq 0.001$) with a rising gradient.
% 		Bright field image overlaid with the trajectory of a migration starved \emph{D. discoideum} cell (green).
% 		\textbf{d.} -------- \textbf{e.} Trajectory plot of cells moving on topotaxis field $r=0.01$. \textbf{f.} Trajectory plot of cells moving on topotaxis field $r=0.01$. \textbf{f.} Drifts ($v_{x,y}$) measured in all live cell experiments (flat,  $r = 0.01$ and  $r = 0.03$) compared. The anisotropicity in pillar positioning was in the $x$ direction for all non-flat experiments. The topotactic drift ($v_{x}$) was found to be significantly higher ($p \leq 0.01$) than noise ($v_{y}$) for all topotaxis assays. Additionally, the topotactic drift was found to increase ($p \leq 0.001$) with a rising gradient.
\label{fig:IVTopotaxis}}
	\end{figure*}

Before investigating the physical origin of topotaxis through multiple \textit{in silico} models, topotactic drift was first experimentally (re)confirmed in square pillar lattices with a continuously changing pillar density. Although earlier work already focused on long-range topotactic drift for amoeboid motility in fields of anisotropically placed cell-sized obstacles, the drift in these experiments was measured on trigonally arranged pillar fields with a  discontinuous gradient \cite{Wondergem2019}, instead of a square lattice with a continuous gradient. The square pillar lattice with a continuous gradient was later introduced in \cite{Schakenraad2020}, and provides a more well-defined approach to studying gradient dependent differences in topotactic response. Contrary to using experimental data from previous experiments, measuring cell migration on the same lattices used for modeling topotaxis ensures a more optimal comparison between experimental and simulated data. Figure~\ref{fig:IVTopotaxis}A,B,C shows SEM-micrographs of the new square PDMS pillar grids used for cell motility experiments here, which were produced using soft photo-lithography techniques similar to earlier work (see \nameref{sec:M&M} for details). The pillars were placed according to Eq.~\ref{eq:pillarcenters}, as introduced in \cite{Schakenraad2020}, and their overall design parameters are shown in Table~\ref{tab:gridpar}.

\iffalse
In earlier work, long-range topotactic drift was measured using anistropically placed cell-sized obstacles in trigonally arranged pillar fields with a  discontinuous gradient \cite{Wondergem2019}.  The pillar gradient changed discontinuously, because the spacing between pillars was increased by \SI{1}{\micro\meter} steps every two or four pillar rows in the direction of the gradient.  as used in this work (see section \nameref{def:pillars}). This type of topotaxis generating lattice was introduced in \cite{Schakenraad2020}, and provides a more well-defined approach to studying gradient dependent differences. Moreover,  \fi

\begin{table}
    \caption{Pillar gradient grid parameter values for the experimental and simulation setup}
    \centering
    \begin{tabular}{c|l|c |c  c} & & Experiment & \multicolumn{2}{c}{CPM}\\
    Symbol & Description  & (\si{\micro\metre})  & (px) & (\si{\micro\metre})  \\
    \hline
    $R$     &  pillar radius & 5 & 10 & 5.3\\
    $h$ & pillar height &18 \textpm 2 & \multicolumn{2}{c}{N/A}\\
    $d$ & pillar center distance at origin & 16.8 & 32 & 16.8 \\
    $d_{\min}$ & pillar center distance left cap & 13.6 & 26 & 13.7 \\
    $d_{\max}$ & pillar center distance right cap & 19.9 & 38 & 20.0 \\
    $s$ & gradient steepness & 0.01, 0.03 & \multicolumn{2}{c}{0.01 - 0.09}\\
    \hline
    \end{tabular}

    \label{tab:gridpar}
\end{table}

\begin{figure*}
    \centering
    \includegraphics[width=\columnwidth]{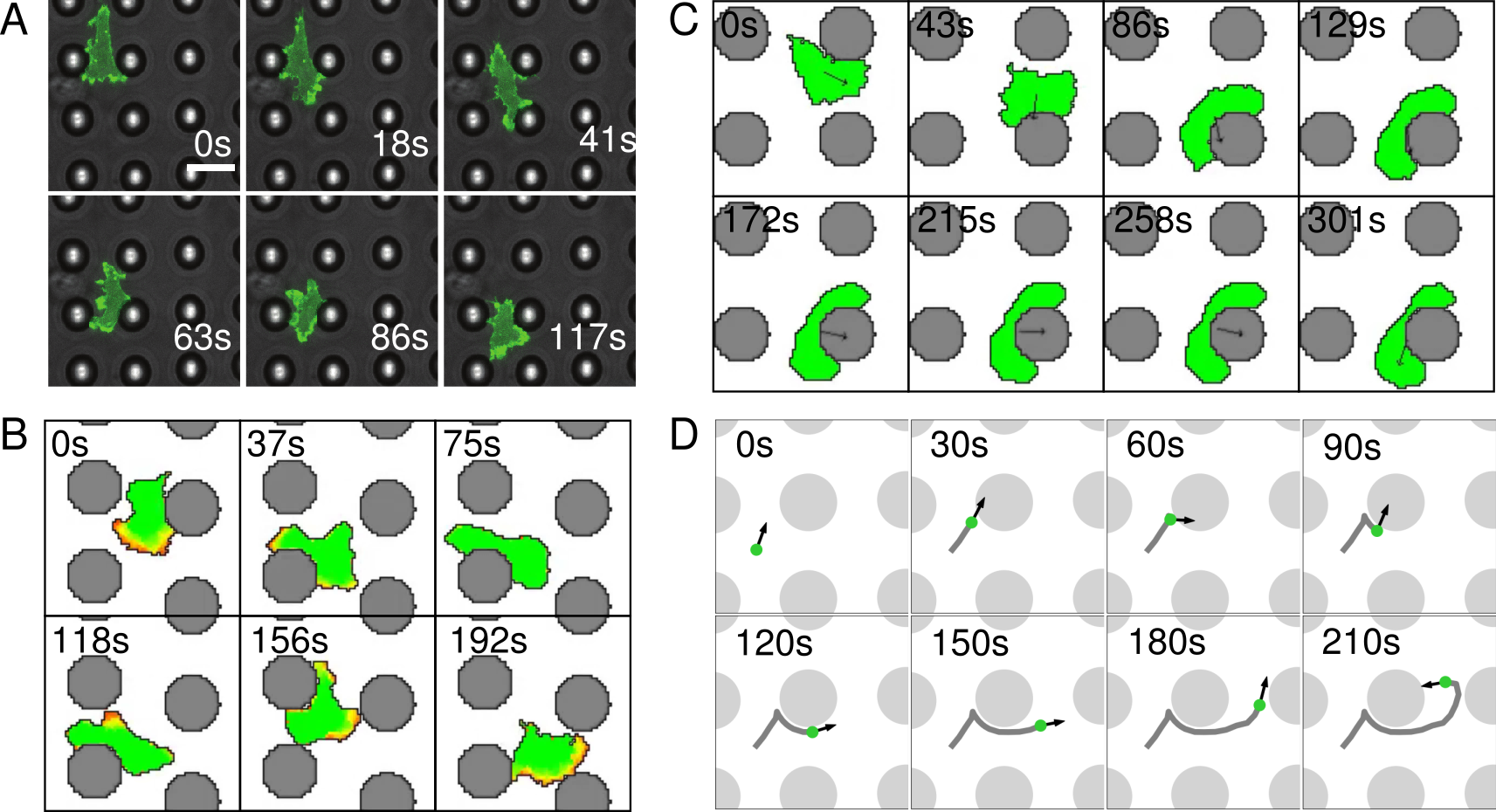}
    \caption{\textbf{Examples of frontal collision with a pillar of \textit{in vitro} starved \textit{D. discoideum} and simulations} (A) Starved \textit{D. discoideum} navigating through a topotaxis assay. The degree of actin polymerization (green) is visualized through fluorescence of LimE-GFP. (B) Act-CPM simulation: Act-levels are colored from red ($Act(\vec{q})=\text{Max}_{Act}$) to green ($Act(\vec{q})=0$). Once the polarized cell hits the pillar, it loses its polarization. A new Act-front appears, the cell polarizes again and moves away from the pillar.
    (C) ABP-CPM simulation: the arrow is the preferential direction vector $\mathbf{p}$ of the cell, starting at its center of mass. Once the cell collides with the pillar, it can only move away once $\mathbf{p}$ points away from the pillar (final frame).
    (D) ABP model simulation: the black arrow is the particle's directional vector $\mathbf{p}$, grey line displays the particle's trajectory. For particles colliding with a pillar, the displacement's component normal to the pillar surface is lost, resulting in motion away from the pillar only when $\mathbf{p}$ has shifted away from the pillar.
    See also \nameref{mov:actvispil}, \nameref{mov:simstarvedpil}.}
    \label{fig:collision}
\end{figure*}

The $s=0.01$ and $s=0.03$ gradient pillar fields (see Fig.~\ref{fig:IVTopotaxis}A,B) were chosen as suitable benchmarks for measuring the topotactic response of starved \textit{D. discoideum} on the new pillar gradients. Measuring topotaxis on steeper gradients ($s>0.03$) was found to be experimentally unfeasible, as these grids produce few and short trajectories in combination with sparse cell seeding, and very gradual gradients ($s \ll 0.01$) are expected to result in low drifts, possibly within experimental error. The cells navigated smoothly between the pillar obstacles (Figures~\ref{fig:IVTopotaxis}D,E and \ref{fig:collision}A, \nameref{mov:actvispil}) during migration experiments (0-3 \si{\hour}). Cell movement was imaged using fluorescent confocal microscopy and tracked using image analysis (Figure~\ref{fig:IVTopotaxis}E, also see \nameref{sec:Imaging_tracking}). The experiments yielded more than a hundred trajectories per pillar gradient, plotted centred in Figure~\ref{fig:IVTopotaxis}F,G. In order to analyze long-range topotaxis of the cells, the drift of the cell center in both the $x$ and $y$ direction ($v_x$ and $v_y$) was calculated from cell tracks in the gradient (Figure~\ref{fig:IVTopotaxis}H). The cells have a significant drift in the positive $x$-direction (i.e. towards sparser pillar densities) for both pillar gradients, when compared to the corresponding $v_y$ in the gradient, as well as compared to movement on flat surfaces (Figure~\ref{fig:IVTopotaxis}H). Moreover, the cells on the steeper gradient ($s=0.03$) showed a significantly larger drift relative to the cells on the shallower ($s=0.01$) gradient.

% \todo[author=Leonie]{Add instantaneous speed to figure, compare with empty field}

The fluorescence time-lapse microscopy of starved \textit{D. discoideum} migrating on flat PDMS served as a control for the topotaxis data, but  also served as the baseline data necessary to empirically calibrate the input parameters required for CPM simulations. The motility of the cells on flat PDMS (Figure~\ref{fig:TuningCPM}, \nameref{mov:actvisstarved}) yielded the base mean cell area (Figure~\ref{fig:TuningCPM}D), mean speed (Figure~\ref{fig:TuningCPM}E), persistence time (Figure~\ref{fig:TuningCPM}G) and diffusion constant (Figure~\ref{fig:TuningCPM}H), summarized in Table~\ref{tab:fits}, necessary as initial input to generate CPM trajectories. In general, the experimental data shows that starved \textit{D. discoideum} cells are highly motile when compared to other cell types \cite{Maiuri2015ActinPersistence}, and move persistent-randomly with a relatively similar speed and persistence time across the cell population for long measurement times \cite{Takagi2008}. The cell area and perimeter are equally similar over time and population, even though the cells continuously grow and retract pseudopods (see Figure \ref{fig:TuningCPM} and \cite{Tweedy2013}).

%\subsection*{Establishing a baseline for \textit{in silico} cell motion}
%
The ABP, Act-CPM and ABP-CPM simulations applied in this study require different sets of input parameters. ABPs are fully defined by a constant speed $v_{0}$ and persistence time $\tau_p$, whereas CPM simulations require an input target cell-area $A_{\sigma}$, cell-perimeter $P_{\sigma}$, and to set the persistence parameters $\lambda$ and $\text{Max}_{Act}$ or $\tau_p$, need an input diffusion constant $D$ and persistence time $\tau_p$ to calibrate motility. Furthermore, the time- and spatial resolution of the CPM system was set by comparing the typical cell area and speed of \textit{D. discoideum} to the time and length scale of MCS simulations (see Table~\ref{tab:fits} for choice). \textit{In silico} cell motion was fit to represent experimental motion by weighing up observables extracted from simulated trajectories to those gathered from experimental trajectories (via the same procedures, i.e. equal frame rate and computational method). The ABP simulations were simply set by using the experimentally determined mean cell speed and persistence time, which resulted in only slightly slower and less diffusive motion (Table~\ref{tab:fits}). In case of CPM-simulations however, the observables were compared through a fitting procedure (see section \nameref{sec:Mod} for details) to obtain a best match to experimental motility. The MCS parameter values used for Act-CPM ($\lambda_\text{Act}$ and $\text{Max}_\text{Act}$) and ABP-CPM ($\lambda_\text{persistence}$ and $\tau_p$) are shown in Table~\ref{tab:par}, with the corresponding motility properties shown in Table~\ref{tab:fits}. See \nameref{mov:simstarvedflat} for example simulations of both systems on a flat substrate. Overall, the data summarized in Table~\ref{tab:fits} indicate that we achieved a close match between simulated and experimental trajectory observables for all three models on flat surfaces.
% \todo[author=Leonie]{Nog een of andere zin dat we denken dat deze fits goed genoeg zijn?}

\begin{table*}
\caption{Cell motility properties for starved \textit{D. discoideum} on a flat PDMS surface, the resulting properties for the ABP simulations, and best fits for the Act-CPM and ABP-CPM.}
\centering
\begin{tabular}{l|llll}
                                                                                         & Experimental                     & ABP                              & Act-CPM                         & ABP-CPM                         \\ \hline
Length equivalent 1px (\si{\micro\metre}) & -                                & -                                & 0.525                           & 0.525                           \\
Time equivalent 1 MCS (\si{s})                                          & -                                & -                                & 0.373                           & 0.574                           \\
Speed (\si{\micro m/s})                                  & 0.197\textpm0.001 & 0.193\textpm0.000 & -                               & -                               \\
Area (\si{\micro m^2})                  & 110.4\textpm45.1  & -                                & -                               & -                               \\
Effective diffusion (\si{\micro m^2/s}) & 1.82\textpm0.68   & 1.77\textpm0.04   & 1.76\textpm0.08  & 1.87\textpm0.08  \\
Persistence time (\si{s})                                               & 91.98\textpm0.98  & 92.0\textpm1.7    & 89.30\textpm2.22 & 89.89\textpm3.33
\end{tabular}

\label{tab:fits}
\end{table*}

\begin{table}
\caption{Parameter values of the best fits for the starved \textit{Dictyostelium} cells for both the Act-CPM and ABP-CPM}
\label{tab:par}
\centering
\begin{tabular}{l|l|l|l}
Parameter                     & Starved & Parameter & Act-CPM \\ \hline
T                                       & 20 & $\lambda_{Act}$ &  129\\
$A_\sigma$                                    & 400 & $\text{Max}_{Act}$    &  37 \\\cline{3-4}
$\lambda_\text{area}$        & 50 &  & ABP-CPM\\\cline{3-4}
$P_\sigma$                              & 313 & $\lambda_\text{persistence}$ & 159\\
$\lambda_\text{perimeter}$   & 1 & $\tau_p$                  & 30\\
$J_\text{cell,medium}$                      & 20 & & \\
\hline
\end{tabular}
\end{table}

\subsection*{Method of persistence affects topotactic capacity of cells}
% For the ABP model, the particle-pillar interaction is described as a frictionless hard wall force that cancels the velocity component normal to the pillar surface whenever the particle would penetrate the pillar. For the CPM simulations, pillars were modeled as immobile obstacles. Cells could not move into a pillar, but could retract from them. Adhesion between cell and pillar can be controlled by the contact energy parameter $J_\text{cell,pillar}$. We assume that cells show no preference for pillars nor medium, so the contact energies between cell-pillar and cell-medium are equal ($J_\text{cell,pillar}=J_\text{cell,medium}=20$), unless stated otherwise.

\begin{figure*}
    \centering
    \includegraphics[width=\textwidth]{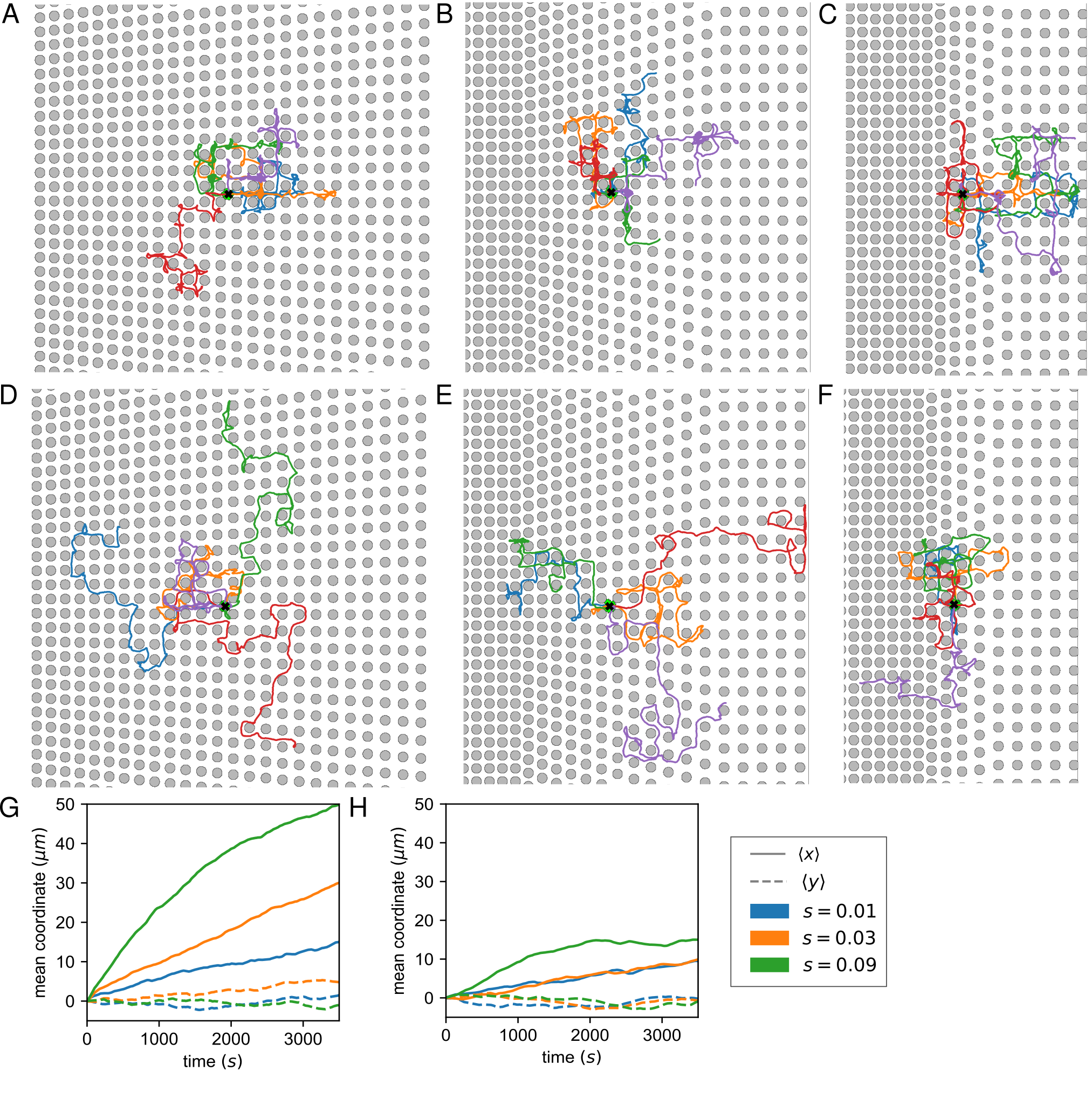}
    \caption{\textbf{Trajectories of starved simulated cells and means of $x$ and $y$ coordinates over time for different gradient steepness parameter ($s$)-values.} \emph{Top row} Act-CPM example trajectories:
    (A) $s=0.01$,
    (B) $s=0.03$,
    (C) $s=0.09$.
    \emph{Middle row} ABP-CPM examples:
    (D) $s=0.01$,
    (E) $s=0.03$,
    (F) $s=0.09$.
    Starting location is marked with a black cross and the depiction of a cell. Each trajectory has its own color.
    \emph{Bottom row} mean $x$ and $y$ coordinates over time for gradient steepness $s$  corresponding to the gradients shown in the rows above.
    (G) Act-CPM.
    (H) ABP-CPM.}
    \label{fig:r_examples}
\end{figure*}

% In contrast to the experiments, repeated simulations were done with single cells only, starting from the same initial position in the center of the field, see Figures~\ref{fig:r_examples},\ref{fig:J_examples},\ref{fig:examples_vegetative_r}.
After confirming topotactic drift in square pillar grids with a continuous spatial gradient, and calibrating the simulations of persistent-random cell motion with experimental data, we turned to an extensive \textit{in silico} investigation of cell-sized obstacle topotaxis. Figure~\ref{fig:collision} summarizes how the various simulated cell bodies respond to obstacles and provides comparison to a representative cell-pillar interaction of starved \textit{D. discoideum}. The amoeba has a single actin front before interacting with a pillar (green, Figure~\ref{fig:collision}A, 0-\SI{18}{\second}), which splits into two upon collision (Figure~\ref{fig:collision}A, 18-\SI{41}{\second}). One of the two newly formed actin fronts becomes the dominant side during the collision, which then guides the entire cell body around the pillar (Figure~\ref{fig:collision}A, 63-\SI{117}{\second}). The Act-CPM cell-pillar interaction follows a similar, CIL-like mechanism of navigating around obstacles (Figure~\ref{fig:collision}B). After colliding with a pillar, the cell temporarily loses actin-polarization (Figure~\ref{fig:collision}B, \SI{75}{\second}), and then repolarizes in a new direction (Figure~\ref{fig:collision}B, 118-\SI{192}{\second}). The ABP-CPM and ABP pillar collisions develop very differently, as these cells or particles always propagate via a self dictated desired direction of motion, therefore, pillar interactions never alter their (inherent) persistence (Figure~\ref{fig:collision}C,D). For these models, the pillars act as areas of excluded volume, unable to alter intrinsic persistence, and only guide the cells around the pillar wall through the velocity component of the cell that does not point directly at the pillar center. The two ABP simulations react differently to the same method of unaltered persistence. The ABP particles always remain in motion, gliding past pillar walls (Figure~\ref{fig:collision}D, 90-\SI{150}{\second}), whereas the inclusion of a cell body for ABP-CPM can lead to prolonged stationary states through cell-pillar embraces.
%The ABP-pillar collisions resolve more quickly than the cell-pillar interactions of the amoeba and CPM models, which typically last in the order of minutes.
% \todo[author=Leonie]{Vergelijken met zelfde soort opname uit experiment}
% \todo[author=Leonie]{A en B omdraaien}

\begin{figure*}
    \centering
    \includegraphics[width=\textwidth]{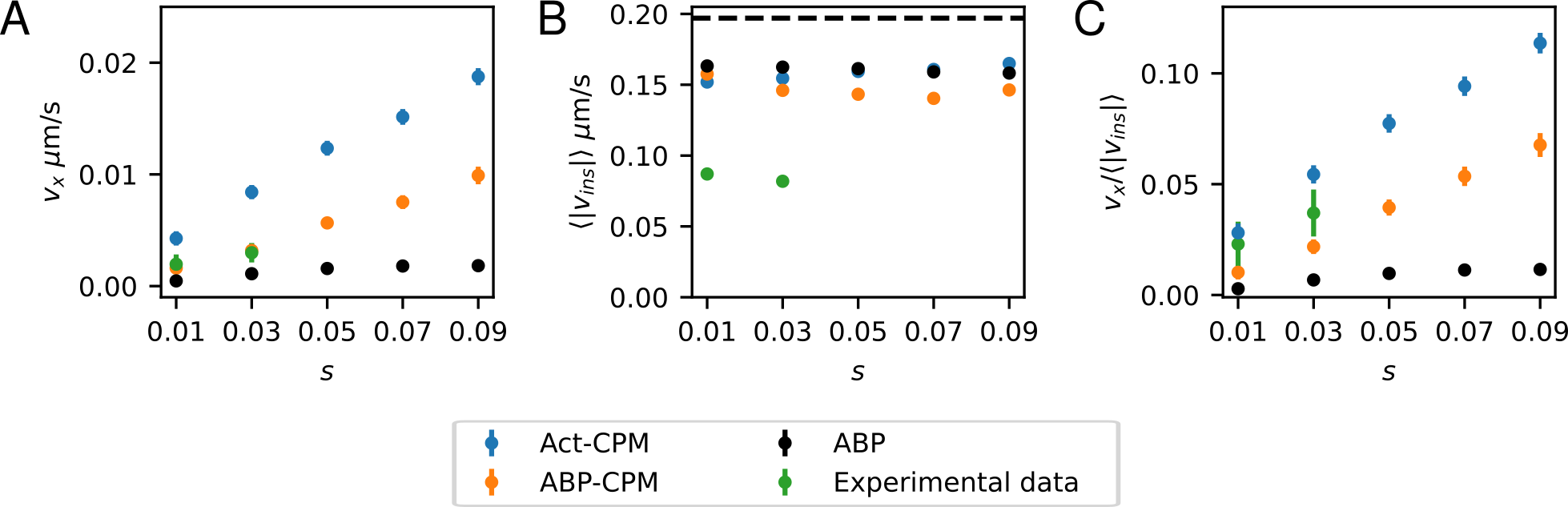}
    \caption{\textbf{Topotactic drift and instantaneous speed of starved parameter set against gradient steepness $s$.} Each orange or blue data point represent the average of a 1000 simulations. Error bars indicate 97.5\% CI.
    (A) Topotactic drift $v_x$.
    (B) Instantaneous speed $\langle | v_{ins}| \rangle $. Dotted line depicts the instantaneous speed of starved cells in absence of pillars.
    (C) Relative topotactic drift $v_x/\langle| v_{ins} |\rangle$).}
    \label{fig:starved_r}
\end{figure*}

% \todo[inline,author=Leonie]{Rewrite definition topotactic velocity: average velocity over time step, correct for initialization, end time with $<5\%$ exited cells}
%We further looked at the behaviour of multiple simulations of the CPMs. Cells move within the pillar grid, but only explore a limited area each (Figure \ref{fig:r_examples}A-F, \nameref{mov:simstarvedpil}).
Example trajectories of ABP- and Act-CPM simulations are shown in Figure \ref{fig:r_examples}A-F (\nameref{mov:simstarvedpil}) for three different gradients. Act-CPM trajectories (Figure~\ref{fig:r_examples}A-C) do not penetrate the dense side of the grid as much as ABP-CPM cells (Figure~\ref{fig:r_examples}D-F). The mean population displacements ($\langle x(t)\rangle$ and $\langle y(t)\rangle $,  Figure~\ref{fig:r_examples}G,H) show that there is drift along the $x$-axis for all gradient steepness ($s$), indicating that the simulated cells perform topotaxis in both models. \iffalse The migratory drift in the $x$-direction ($v_{x}$) for the experimentally measured topotaxis (Figure~\ref{fig:IVTopotaxis}H), the ABP model and both CPMs with the starved parameter sets are shown in Figure \ref{fig:starved_r}A as a function of $s$.\fi The drift velocity in the direction of the gradient ($v_{x}$, Figure~\ref{fig:starved_r}A) increases as a function of steepness for the trajectories of both the experimental data and all models. The models have a mean drift proportional to the gradient steepness, as described for ABPs by Schakenraad et al. \cite{Schakenraad2020}, but there is a clear difference between the ABP and CPM simulations: CPMs show a higher increase in drift than the ABP model, and furthermore, the Act-CPM is most effective at performing toptaxis. When directly comparing the topotactic drift between simulations and \textit{D. discoideum}, the ABP-CPM predictions are closest to the experimental data, with Act-CPM drifting more, and the ABP model underestimating amoeboid topotaxis. In case of ABPs, there is a small effect of particle size on the topotactic drift ($R_p\in\{0,1,1.8\}$, see Figure~\ref{fig:abp_r} for more details).

% \begin{table}[]
% \caption{Instantaneous speed $\langle |v_{ins}|\rangle$ in \si{\micro m/s} on empty field and within pillar grid for starved \textit{D. dictyostelium} cells, starved active reorientation model simulations and starved passive reorientation model simulations. Given error is 97.5\% confidence interval.}
% \begin{tabular}{l||c|c|c}
%  Field           &  Experimental & Act-CPM  & ABP-CPM \\
%  \hline
% Empty field   & 0.197\textpm 0.00139 & 0.198\textpm 0.00285 & 0.197\textpm 0.001331\\
% $s=0.01$      & 0.087\textpm 0.00080 & 0.152\textpm 0.00046 & 0.156\textpm0.000264\\
% $s=0.03$      & 0.082\textpm 0.00082 & 0.156\textpm 0.00047 & 0.148\textpm 0.000274
% \end{tabular}
% \label{tab:speeddrop}
% \end{table}

% \todo[author=Leonie]{Ophelderen en vanuit perspectief exp. data benaderen.}
Directly comparing the unnormalized topotactic drift $v_{x}$, assumes that the base motile response across all the investigated systems does not differ in a gradient versus on a flat substrate, or, for that matter, as a function of gradient steepness. However, earlier work by Schakenraad et al. suggest that this might not be the case. The mean instantaneous velocity $\langle|v_{ins}|\rangle$, an often used indicator of base cellular motility \cite{Selmeczi2005,Cherstvy2018}, is affected by pillar distance and since cells in steeper gradient encounter different pillar distance more quickly, their instantaneous velocity can vary across different gradients. The $\langle|v_{ins}|\rangle$ is shown in Figure~\ref{fig:starved_r}B for the experimental- and simulated data, and develops differently for each model. The mean speed of ABPs, \textit{D. discoideum} and ABP-CPM cells drop with increasing gradient steepness, although at different rates, whereas the Act-CPM trajectories gain motility as a function of steepness.  \iffalse As Figure~\ref{fig:collision} shows, \textit{D. discoideum} and simulated cells collide with pillars. This decreases their speed. \fi Overall, the instantaneous velocities always drop when compared to movement on flat substrates, but \textit{D. discoideum} features a larger decrease in speed than the simulated cells. To properly compare the topotactic response across all the investigated systems, the drift must be normalized to the mean instantaneous velocity to account for the overall, dissimilar drop in motility on pillar substrates, and for any differences in speed response to varying the gradient steepness. Normalizing the topotactic drift ($v_x/\langle|v_{ins}|\rangle$, Figure~\ref{fig:starved_r}C) increases the distinction between ABP and CPM-cells, highlighting the influence of cellular volume on cell-sized obstacle topotaxis, and also positions the experimental topotactic response  between the two CPM simulations.

Summarizing the \textit{in silico} results so far, the inclusion of a dynamic cell body in the CP-models clearly influences the overall topotactic response, and provides a more accurate estimation of the topotactic drift measured for \textit{D. discoideum} \textit{in vitro} than ABP simulations. Especially the comparison between ABP and ABP-CPM,  highlights this finding, as the two models share the exact same method of generating persistence, but the pillar collisions ensue differently for ABP-CPM through the inclusion of a dynamic cell body, which seems to temporarily decrease motility during pillar interactions (Figure~\ref{fig:collision}). In case of Act-CPM, the CIL-like pillar interaction may disrupt persistence (or motility) most, which seems to translate into the strongest topotactic drift. If cell-pillar interactions indeed cause different decreases in motility or persistence, this is important, as obstacle induced disruptions of motion can lead to the emergence of topotaxis through the anisotropy in pillar density along the direction of the gradient. A positionally dependent difference in persistence leads to a net population drift in the direction of the gradient, see \cite{Novikova2017,Schakenraad2020}. In the next section we will investigate if the model with the strongest topotactic reponse, in this case Act-CPM cells, is indeed accompanied by the largest difference in obstacle induced disruption of persistence or motility along the gradient axis.

\subsection*{Cell-pillar interactions are most disruptive for Act-CPM persistence}

\begin{figure*}
    \centering
    \includegraphics[scale=1]{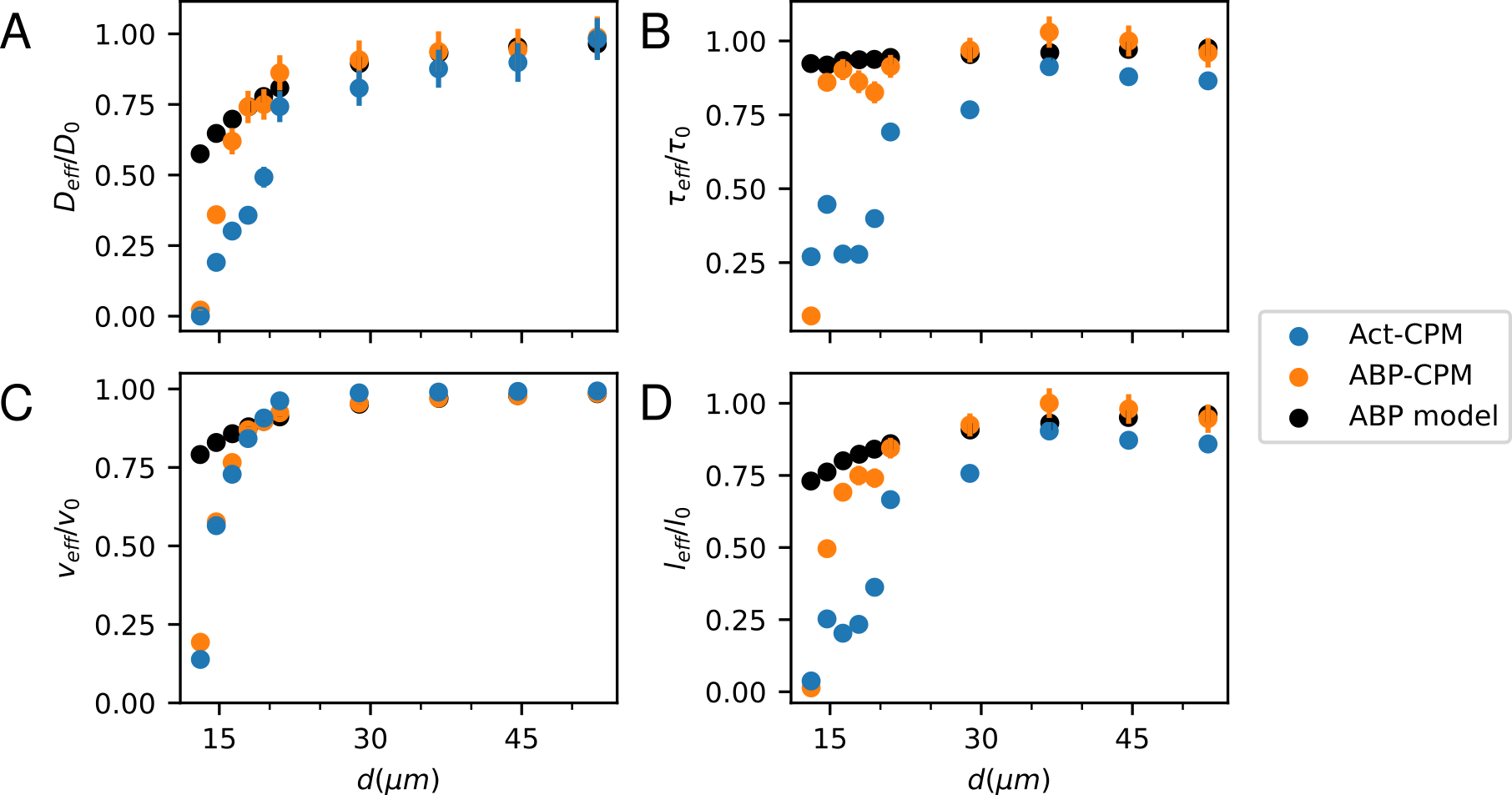}
    \caption{\textbf{Normalized effective parameters of the persistent random walk in regular pillar grids for the Act-CPM, ABP-CPM and ABP model with the starved parameter settings.}
    (A) Effective diffusion coefficient $D_{\text{eff}}$ normalized by the diffusion coefficient of starved simulations on an empty field as a function of the distance $d$ between pillar centers.
    (B) Effective persistence time $\tau_{\text{eff}}$ normalized by the persistence time of starved simulations on an empty field as a function of $d$.
    (C) Effective instantaneous speed $v_{\text{eff}}$ normalized by the instantaneous speed of starved simulations on an empty field as a function of $d$.
    (D) Effective persistence length $l_{\text{eff}}=v_{\text{eff}}\tau_{\text{eff}}$ normalized by the persistence length of starved simulations on an empty field as a function of $d$. Error bars indicate 97.5\% CI.}
    \label{fig:reg}
\end{figure*}

Inspired by Novikova et al. \cite{Novikova2017}, Schakenraad et al. \cite{Schakenraad2020} attributed topotaxis to changes in effective persistence length $l_{\text{eff}}=v_{\text{eff}} \cdot \tau_{\text{eff}}$ as a function of position on the gradient axis. They argued that the increase in pillar density in the direction of the gradient leads to longer persistence lengths during movement, and therefore causes drift in the direction of the gradient. Here, we will extend the results of Schakenraad et al., by calculating the same observables of persistent-random motion for the Act-CPM and ABP-CPM systems that were previously determined for ABPs. First, the diffusion constant $D_{\text{eff}}$ was calculated by a fit of the second moment of Brownian motion in two-dimensions (Eq.~\ref{eq:brownian}) to the ensemble-averaged MSD (EAMSD, Equation \ref{eq:EAMSD}) of the trajectories at long lag-times,i.e., during the diffusive regime of persistent-random motion. Subsequently, the persistence time $\tau_{\text{eff}}$ was determined by fitting the full EAMSD to the F\"urth's equation for persistent-random motion (Eq.~\ref{eq:furth} for the ABP model, and corrected for noise by an extra translational diffusion term for the CPMs, Eq.~\ref{eq:furthtransdiff}, see Methods). As a final step, the instantaneous velocities $v_{\text{eff}}$ were measured by computing $\langle |v_{ins}|\rangle$ for each pillar configuration, and subsequently used to determine the associated persistence lengths ($l_{\text{eff}}$). In general, the data was well fit by the theoretical curves for F\"urth and Brownian MSDs, with the exception of the recorded trajectories on the pillar grids with $d=$~\SI{13.1}{\micro\metre}, where the cell bodies were mostly stuck at their starting unit cell, leading to strong subdiffusive behavior.

% Link naar eerdere observaties: snelheden liggen weinig uit elkaar
Figure \ref{fig:reg} shows the effective motility parameters, or key observables for persistent-random motion as a function of pillar distance, of both the CPMs and ABP model, normalized to the motility of each system on a flat substrate. The effective diffusion coefficient $D_{\text{eff}}$, persistence time $\tau_{\text{eff}}$, instantaneous speed $v_\text{eff}$ and persistence length $l_{\text{eff}}$ all increase as a function of increasing pillar distance, and saturate to the corresponding values measured on flat substrates ($D_{0}$, $\tau_{0}$, $v_{0}$ and $l_{0}$). The speed $v_\text{eff}$ is a measure of the magnitude of the velocity vector, and thereby motility, where the persistence time $\tau_{\text{eff}}$ contains information on both the direction and rate of motion, hence it functions as an indicator for general persistence \cite{Stokes1991}. The effective speed of CPM cells decreases faster than ABPs for small pillar spacings, but does not exhibit a very pronounced difference between the Act-CPM and ABP-CPM. On the contrary, the effective persistence time of the two types of CPM is very dissimilar\textemdash~the Act-CPM simulations have much shorter persistence times than ABP-CPM tracks at small pillar spacings. Hence, the ABP-CPM system is mainly affected by a decrease in motility, while Act-CPM is affected by both a decrease in persistence and the overall decease in motility.

Both the deformable cell body and the additional disruption of persistence are key ingredients in enlisting a stronger topotaxtic response than measured for active particles. So far, the CPM models were calibrated to starved \textit{D. discoideum} movement, which typically has a persistence length larger than the cell body. However, when the persistence length of the system is shortened, one expects the contribution of obstacle induced disruption of persistence to the topotactic drift to decrease, and both types of CPM motility likely result in the same drift. In order to test this hypothesis, we calibrated the CPM models to represent vegetative \textit{D. discoideum}, which have a much shorter persistence length than their starved counterparts (see Appendix \ref{sec:vegetative} in \nameref{supfile} for details). Both CPMs predict some topotactic drift (Figure~\ref{fig:vegetative_r}), more than the ABP model, but five times smaller than for the starved simulations. More interestingly however, there was no clear difference between the two CPM simulations. The cell-simulations in regularly spaced grids show that there is indeed a much smaller difference in effective persistence times as a function of pillar spacing, especially for pillar fields spaced apart \SI{16.3}{\micro\meter} or more (Figure~\ref{fig:reg_veg}). Summarizing, the deformable cell body of CPMs impairs cell speed more during cell-pillar interactions when compared to hard-particles, like ABPs, and enhances topotactic drift. Secondly, methods of motility that are more susceptible to obstacle induced disruption of persistence lead to even higher topotactic response, as the comparison of Act-CPM and ABP-CPM shows. Finally, setting simulations of vegetative and starved cells side by side, shows that for persistence lengths much shorter than the average cell diameter, any differences in persistence generation hardly contribute to the net topotactic drift, except for the narrowest of pore sizes.

\subsection*{Pillar adhesion influences cell motility and topotactic response}

Up to this point, cell-medium and cell-pillar interactions were treated equally in the CPM simulations as the adhesion energies of both components of the cell environment were chosen the same ($J_\text{cell,pillar}=J_\text{cell,medium}$). However, this might not properly reflect the response of starved \textit{D. discoideum} to the pillar walls. For one, the live-cell experiments led to a much larger drop in cell speed when going from featureless surfaces to pillar assays, when compared to the simulations (Figure \ref{fig:starved_r}B). Moreover, previous work \cite{Wondergem2019} has shown that the amoeba occasionally explore the vertical pillar walls, which leads to temporary stationary states in persistent cell motion when next to pillars, not unlike the Act-CPM CIL-like interaction between cells and pillars. Drops in speed, lingering behavior next to obstacles, or pillar guidance may be simulated by varying the adhesion energy of pillars relative to the surrounding flat medium $J_\text{pillar,medium} \neq J_\text{cell,medium}$. During all simulations the cell-medium adhesion energy was kept positive and constant ($J_\text{cell,medium} = 20$, see Table \ref{tab:par}). Lowering the cell-pillar adhesion energy relative to the cell-medium energy ($J_\text{pillar,medium} < J_\text{cell,medium}$), promotes occupying lattice sites next to pillars, and thereby encourages cell-pillar adhesion, while discouraging movement into empty medium. Conversely, increasing cell-pillar adhesion energy impedes occupying lattice sites next to pillars, discouraging cell-pillar adhesion and encouraging movement away from pillars into empty medium.

\begin{figure*}
    \centering
    \includegraphics[width=\textwidth]{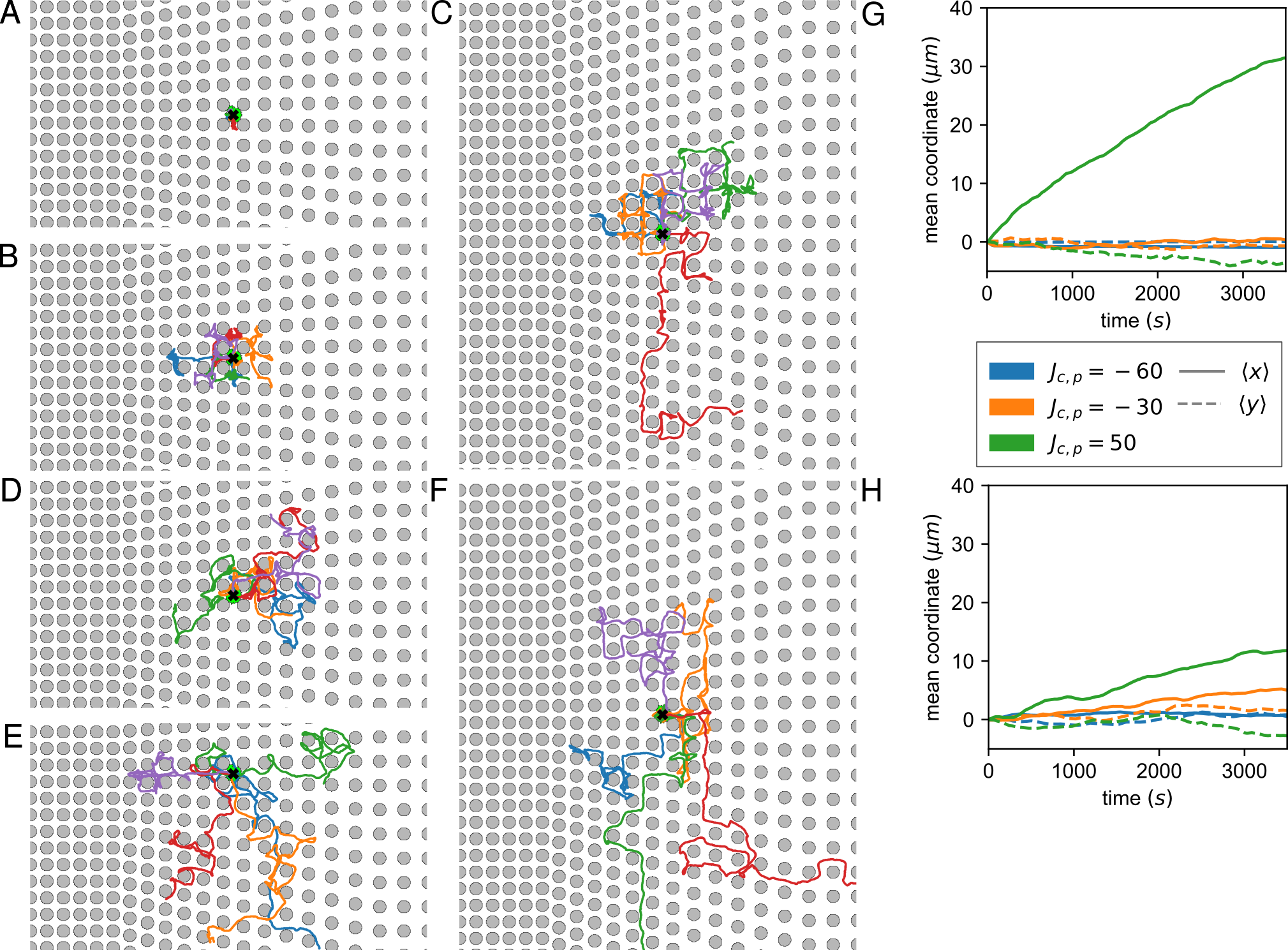}
    \caption{\textbf{Trajectories  of  starved  simulated  cells and mean $x$ and $y$ coordinates over time for  different  pillar adhesion energy $J_\text{cell,pillar}$.}
    \emph{Top row} Act-CPM example trajectories: (A) $J_\text{cell,pillar}= -60$, (B) $J_\text{cell,pillar}=-30$, (C) $J_\text{cell,pillar}=50$.
    \emph{Bottom row} ABP-CPM example trajectories: (D) $J_\text{cell,pillar}= -60$, (E) $J_\text{cell,pillar}=-30$, (F) $J_\text{cell,pillar}=50$.
    Starting location is marked with a black cross and the depiction of a cell. Each trajectory has its own color.
    \emph{Right column} mean $x$ and $y$ coordinates over time of a 1000 independent simulations for (G) Act-CPM and (H) ABP-CPM.}
    \label{fig:J_examples}
\end{figure*}

To investigate the effect of modifying the adhesion energy of cell-pillars relative to the surrounding medium, we varied the adhesion parameter $ -60 < J_\text{cell,pillar} < 50$ around our initial choice $J_\text{cell,medium}=J_\text{cell,pillar}$ in a topotaxis grid with gradient steepness $s=0.03$. Examples of simulated cell tracks for more adhesive pillars $J_\text{cell,pillar} < 0$ and repelling pillars $J_\text{cell,pillar} > J_\text{cell,medium}$ are shown in Figure~\ref{fig:J_examples}. Clearly, decreasing $J_\text{cell,pillar}$ (Figure~\ref{fig:J_examples}A,B,D,E) reduces overall cell displacement in the pillar field, more so for Act-CPM cells than ABP-CPM cells, whereas increasing $J_\text{cell,pillar}$ (Figure~\ref{fig:J_examples}C,F) does not seem to affect cell displacement at first sight (compared to $J_\text{pillar,medium} = J_\text{cell,medium}$, see  Figure~\ref{fig:r_examples}B,E). The mean coordinates over time (Figure~\ref{fig:J_examples}G,H) show that the Act-CPM loses its topotactic drift for low $J_\text{cell,pillar}$, but not for high $J_\text{cell,pillar}$. The ABP-CPM is less affected by high pillar adhesiveness, as it still displays topotactic drift at $J_\text{cell,pillar}=-30$.

% A large distinction between point particle and cells, is that cells interact with their environment with their membranes. One important aspect regarding the cell membrane is its adhesiveness to other substrates.
% To test whether the adhesiveness of cells to pillars affects topotaxis, we simulated cell movement in a grid with gradient $s=0.03$ for different adhesion energies. We let $J_\text{cell,pillar}$ range from $-60$ to $60$, where $J_\text{cell,pillar}=20$ is neutral with respect to $J_\text{cell,medium}=20$.

% There is an effect of decreasing the adhesion energy $J_\text{cell,pillar}$ to negative values, which decreases the topotactic drift. For very negative energy differences, actively reorienting cells eventually do topotaxis in the other direction.
% Increasing $J_\text{cell,pillar}$ with respect to $J_\text{cell,medium}$ does not seem to have an effect on $v_x$, suggesting that the topotactic drift saturates with increasing $J_\text{cell,pillar}$. Again, we see a difference in the topotactic drift between the two persistence methods (Figure~\ref{fig:starved_J}(a)).

\begin{figure*}
    \centering
    \includegraphics[width=\textwidth]{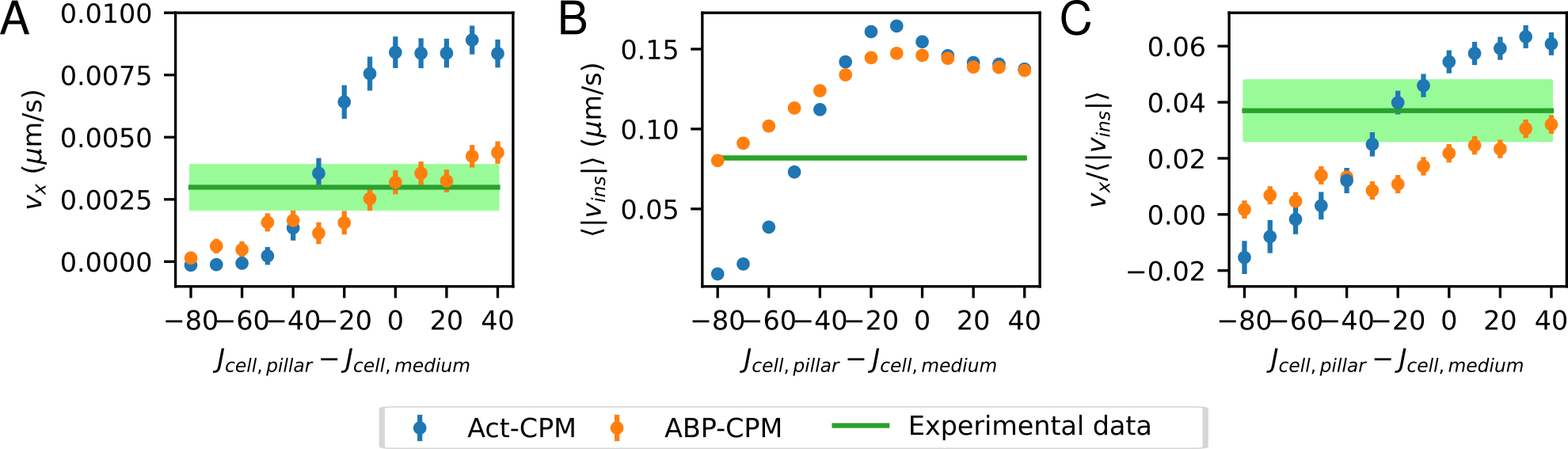}
    \caption{\textbf{Topotactic drift and instantaneous speed of starved parameter set against adhesion preference $J_\text{cell,pillar}-J_\text{cell,medium}$.} For all points $J_{\text{cell, medium}}=20$.
    Each data point represent the average of a 1000 simulations, error bars indicate 97.5\% confidence interval. Line with shaded area represents the experimental data with 97.5\% confidence interval.
    (A) Topotactic drift $v_x$.
    (B) Instantaneous speed $\langle |v_{ins}|\rangle$.
    (C) Relative topotactic drift $v_x/\langle |v_{ins}|\rangle$.}
    \label{fig:starved_J}
\end{figure*}

% sigmoidal velocity with peak in middle for active, passive shows less explicit peak in speed
% Sigmoidal for active, linear trend for passive
To further investigate topotaxis for different adhesion energies of pillars, we quantified the topotactic drift $v_x$ and instantaneous velocity $\langle v_{\text{in}} \rangle$ as a function of the difference in adhesive energy between pillars and the medium ($J_\text{cell,pillar}-J_\text{cell,medium}$). The topotactic drift of Act-CPM cells has a sigmoidal relation to $J_\text{cell,pillar}-J_\text{cell,medium}$, whereas the drift $v_x$ of ABP-CPM increases approximately linearly as a function of $J_\text{cell,pillar}$. Where the drift of the ABP-CPM population seems to keep increasing as a function of adhesion energy, even beyond its neutral value ($J_\text{cell,pillar}-J_\text{cell,medium} = 0$), the drift of Act-CPM trajectories quickly plateaus when $J_\text{cell,pillar}>J_\text{cell,medium}$. Furthermore, high cell-pillar attraction ($J_\text{cell,pillar}\ll 0$) has a large effect on the overall motility of Act-CPM cells, as both drift and instantaneous velocity quickly drop to zero for negative adhesion energies (Figure~\ref{fig:starved_J}B). Hence, lowering the adhesion energy between cells and pillars can be utilized to model the stationary states in 2D-projected \textit{in vitro} cell trajectories, and accompanying drop in speed, that occurs when \textit{D. discoideum} explore the vertical pillar walls. The ABP-CPM response in speed is similar to that of Act-CPM, but the cells lose less motility overall at very low cell-pillar adhesion energies, both models seem to share an optimum at $J_\text{cell,pillar}=10$.
% Relative shows two linear trends, with active saturating.

% Herschrijven
The initial aim of varying the cell-pillar adhesion energy was to gauge the effect of the parameter on the strength of topotactic response of the two CPM simulations. In short, decreasing cell-pillar energy negatively affects topotactic drift, and an increase of energy beyond $J_{\text{cell}}$ does not affect Act-CPM trajectories, but does enhance the Act-ABP drift. As a final step, we compare the observables of the \textit{D. disocideum} trajectories with the data for the range of adhesion energies investigated (shown in green in Figure~\ref{fig:starved_J}A-C). The topotactic drift $v_x$ and speed $\langle |v_{ins}|\rangle$ of the models match the experimental data for different adhesion energies, but never at the same energy for both observables (e.g. at $J_\text{cell,pillar} = -50$ for $v_x$ and  $J_\text{cell,pillar} = -70$ for $\langle |v_{ins}|\rangle$ for Act-CPM). The normalized drift $v_{x}/\langle v_{\text{in}} \rangle$, which scales the topotactic response to the base motility of the system, matches he Act-CPM at $J_\text{cell,pillar}-J_\text{cell,medium}=-20$, where the ABP-CPM simulations never provide a good fit. Hence, the relative topotactic drift can be approximated using Act-CPM simulations, although it overestimates the base motility of the live-cell data. Overall, the cell-pillar adhesion energy alters cell speed, but lowering speed comes at a cost of decreased topotactic drift, and the adhesion energies at which the experimental values match the observables $v_x$, $|v_{ins}|$ or $v_x/|v_{ins}|$ are very different between the two CPM simulations.

%Comparing the results of drift and instantaneous speed to the experimental data, we can make two observations. First is that the $J_\text{cell,pillar}-J_\text{cell,medium}$ values at which the model predictions of $v_x$ match to the $v_x$ of \textit{D. discoideum} cells differ for both models: $J_\text{cell,pillar}-J_\text{cell,medium}=-30$ for the Act-CPM and $J_\text{cell,pillar}-J_\text{cell,medium}\in[-10,20]$ for the ABP-CPM. Second is that for these values, $\langle |v_{ins}|\rangle$ is still higher for the simulated data than the experimental data, resulting in underestimation of the relative topotactic velocity. The values of $J_\text{cell,pillar}$ for which the instantaneous speeds are indeed similar between model and experiment are $J_\text{cell,pillar}-J_\text{cell,medium}=-80$

%\subsection*{Persistence method does not influence topotaxis for less persistent motion}
%We also predicted topotaxis for a less persistent type of motion, based on the motility of vegetative \textit{D. discoideum}. Both CPMs predict a small topotactic drift, more than the ABP model (Figure~\ref{fig:vegetative_r}). However, the relative topotactic velocity was two to five times smaller than for the starved simulations, and moreover, there was no clear difference between the two persistence methods. In the regularly spaced grids, the effective persistence time and length were only lower for the Act-CPM than for the ABP-CPM at $d\leq$\SI{19.4}{\micro\metre} (Figure~\ref{fig:reg_veg}). For more details, see Appendix \ref{sec:vegetative} in \nameref{supfile}.

\section*{Discussion and Conclusion}

%korte samenvatting van wat we gedaan en gevonden hebben
In this study, topotactic drift was measured for starved \textit{Dictyostelium discoideum} in square lattices of cell-sized pillar-obstacles with continuously changing gradients in pillar density. To gain further insight into the underlying physical principles that yields this kind of long-range topotactic drift, the experimental results were followed up by three types of cell simulations in the same anisotropic pillar fields. Topotaxis was simulated using two different Cellular Potts models, and in order to compare to previous work, were accompanied by Active Brownian particle simulations with the same parameters. The two CPMs modeled persistent-random cell motion through two distinct methods of persistence: the Act model with emergent Contact-Inhibition of Locomotion, and an ABP-CPM hybrid, that leaves pillar obstacles without the capacity to alter the inherent persistence of the cells. The motion of simulated cells was matched to the type of persistent-random motion of \textit{D. discoideum} by fitting the model parameters to the experimental results on flat PDMS, providing a way to perform order of magnitude comparisons between live-cell and simulated data. The CPM-cells had topotactic drifts on par with the live-cell data, contrary to ABP simulations, which have a much lower topotaxis efficiency. Generally, both the experiments and simulations showed that rising pillar-gradient steepness linearly increases topotactic drift, although there are large differences in the origin and strength of the topotactic response of the three models.

\subsection*{Persistence driven cellular topotaxis is primarily affected by changes in persistence time}

Active Brownian motion on a plane experiencing increasing and decreasing persistence length in opposing directions was previously shown to result in a net drift \cite{Novikova2017}. As undirected cell motility is similar to active Brownian motion at long measurement times \cite{Selmeczi2008CellReview,Campos2010}, positionally dependent differences in persistence length are proposed to be a key ingredient of long-range topotaxis \cite{Schakenraad2020}. In topotaxis simulations by Schakenraad et al., ABPs glide past pillar walls and lose the part of the velocity vector that points in the direction of the pillar center, and thereby some persistence length. Therefore, a gradient in pillar density leads to either larger or smaller losses of velocity depending on gradient position, which in turn, leads to a net topotactic drift. Cells are not spherical particles however, and obstacle induced ABP topotaxis was shown to be accompanied by a much lower normalized drift than measured for highly motile amoeba in anistropic cell-size pillar fields \cite{Wondergem2019}. The difference between both systems is likely caused by the capacity of cell-pillar collisions to influence the intrinsic persistence of the cells, whereas ABPs are merely deflected by pillar walls, always maintaining their inherent constant speed and persistence.

We investigated the influence of including a deformable cell body and changing the method of motility on obstacle induced changes in the persistence of cell motion. Each of these two factors generally exacerbate changes in persistence length, and thereby topotactic drift, for persistent-random cell motion through a pillar gradient, but affect the speed or persistence time differently. Specifically, comparing the ABP and ABP-CPM shows that the inclusion of a dynamic cell volume alone is not enough to substantially change persistence time as a function of pillar distance. However, including cell volume does cause a drop in effective speed through the more complex pillar interactions, like the cell body embrace of pillars. Moving on to the second factor, when obstacle collisions are allowed to alter incident persistence through using different methods of motility, it does lead to substantial changes in persistence time, as contrasting the Act-CPM and ABP-CPM simulations shows. The Act-CPM contact inhibition of locomotion cell-pillar interactions lead to a large loss of incident cell polarization, which allows for sharper turns upon collision than the ABP-CPM, which stays more in line with the incident direction of motion. The different methods of motility lead to a large difference in persistence time between the two models, although interestingly, the speed measured during the Act- and ABP-CPM simulations is very similar. The difference in persistence time between the CPMs disappears when the persistence length is decreased beneath the cell diameter (realized by tuning simulations to vegetative \textit{D. discoideum} instead), and the additional contribution of the method of persistence vanishes from the topotactic drift.

The reorientation of cells upon obstacle collision has also been subject of other studies \cite{Vallotton2005TrackingFront,Nishimura2012Non-BrownianLocomotion,Campbell2018AObstacles}. In an experiment with fish keratocytes, actin flow was disturbed upon collision with an obstacle, making the keratocyte change direction \cite{Vallotton2005TrackingFront}. Modeling studies also indicate that intracellular molecular dynamics are important in cell reorientation upon collision. In Nishimura et al. \cite{Nishimura2012Non-BrownianLocomotion} the change in actin retrograde flow plays an important role in the formation of a new protrusion after collision with a wall, allowing the cell to move away from the wall. In Campbell et al. \cite{Campbell2018AObstacles} many different cell behaviours after collision were observed in a immersive-boundary method with reaction-diffusion equations on the cell surface. They observed 1) `freezing' when the cell did not have new pseudopods for a short while after collision, 2) `doubling back' when the cell returned in the direction it came from and 3) `tug of war' when multiple new pseudopods competed for leading the cell's new direction.

% Storm's model lijkt op run-and-tumble in E. coli. Zie bv. \url{https://jb.asm.org/content/jb/172/2/1099.full.pdf}. Zoek even de initiele referentie erbij.

%Verschil tussen Act en vector model illustreert dat nog meer processen dan enkel de verandering van effectieve persistentie topotaxis kunnen beinvloeden.

%Wat verklaart mogelijk het verschil tussen model en experiment?

\subsection*{Comparing cells on silicon and \textit{in silico}}

To achieve CPM simulations close to the experimental data, we explored if cell-pillar adhesiveness is at the root of the discrepancy in instantaneous speed between the experiment and model and if it can be exploited to more accurately model the measured ratio between base motility (speed) and topotactic drift of the experimental data. The instantaneous speed of the simulated cells in both CPMs could be lowered by decreasing the cell-pillar adhesion energy, i.e., by increasing cell-pillar attraction. However, the motility of Act-CPM cells decreases much quicker than in case of ABP-CPM, where the unaltered base persistence is capable of breaking the cell free from pillar contact. Cell displacement decreases at low adhesion energy leading to a reduced overall topotactic response in both simulations. Conversely, the simulations respond differently to increasing cell-pillar adhesion energy beyond that of the medium, as Act-CPM topotactic drift quickly plateaus to a constant value and ABP-CPM keeps rising at increased energy. It was possible to obtain the same normalized drift as the experimental data by varying the adhesion energy of Act-CPM simulations, however, the concurrence is reached at a slightly different magnitude of the instantaneous velocity and drift.

Although our primary aim was identifying the key ingredients that uplift ABP topotaxis to cellular topotaxis in cell-sized pillar fields, the method of \textit{in vitro} inspired CPM simulations applied here also has a potential to help discover useful obstacle assays in the future. By first measuring cell motion experimentally on a flat surface, then fitting CPM model parameters to match that motion, and finally predicting the response of cells to an obstacle configuration, the method can serve to predict the response of other cells to different obstacle designs. Using simulations instead of gathering data on \textit{in vitro} assays, can speed up the design processes of new assays that aim to guide cell motion by obstacles or obstacle adhesiveness in order to, for example, achieve cell sorting or promote tissue regeneration.

In setting up such simulations, one should strive to minimize any differences between the computational and live-cell results. Despite the fact that including a cell body, alterable persistence and pillar adhesion energy delivered the same normalized topotactic drift, our simulations were unable to reproduce the general loss of motility measured for the amoeba. The simulations likely do not fully capture the dimensionality and chemotactic intricacies of the \textit{in vitro} assay. Experimental observation show that \textit{D. discoideum} cells sometimes crawl up a pillar, mainly displacing their center of mass vertically. They can also move through narrower pillar distances than our CPM simulations by spreading their volume vertically. Although vertical displacement is not modeled directly in the CPM, the weight of the area and perimeter constraint can be seen as a measure of 3D flexibility. Starved \textit{Dictyostelium discoideum} secrete and degrade cAMP \cite{Masaki2013}, which acts as a chemo-attractant and promotes chemokinesis. Obstacle configurations may non-trivially influence local cAMP concentrations, for example between pillar domains. Studies have shown that  chemotaxis can assist in guiding cell movement around large obstacles \cite{Grima2007DirectedObstacles}, or through mazes \cite{Scherber2012}. Furthmore, Self-secreted chemoattractants can also trap cells within containing environments such as dead ends in mazes \cite{Marsden2014a}. The effect of chemical sensing in a field with a high density of obstacles as well as its effect in topotaxis are still open for study.

\section*{Author Contributions}
L.v.S, J.A.J.W, K.S, D.H, and R.M.H.M designed the research. L.v.S performed simulations. J.A.J.W. performed experiments. L.v.S and J.A.J.W analyzed the data and wrote the manuscript. D.H. and R.M.H.M critically revised
the manuscript.

\section*{Acknowledgments}
We thank SURFsara for the support and computing time in using the Lisa cluster computer. J.A.J.W and D.H acknowledge the Fraunhofer Society for the Fraunhofer Attract grant “3DNanoCell” for partly funding this work and thank Dr. Günther Gerisch (Max Planck Institute of Biochemistry, Martinsried, Germany) for providing axenic \textit{D. discoideum (Ax2)} with both free GFP and limGFP in lim0 insertion. R.M.H.M. was funded by the Nederlandse Organisatie voor Wetenschappelijk Onderzoek grant NWO/ENW-VICI 865.17.004.

\section*{Supporting Material}

\paragraph{Supporting File 1}\label{supfile} Figures \ref{fig:TuningCPM}-\ref{fig:reg_abp}, Appendix A, and Movies.

    \paragraph{Supporting File 2} Implementation of the CPMs.

% Uncomment if using bibtex (default)
% \include{bibliography_short}
\bibliographystyle{biophysj}
\bibliography{BPJ_mainv0p2_biber}

\renewcommand{\thefigure}{S\arabic{figure}}
\renewcommand{\thetable}{S\arabic{table}}
\setcounter{figure}{0}
\setcounter{table}{0}

\clearpage

\section*{Supplementary Figures}

\begin{figure}[h]
		\includegraphics[width=0.9\textwidth]{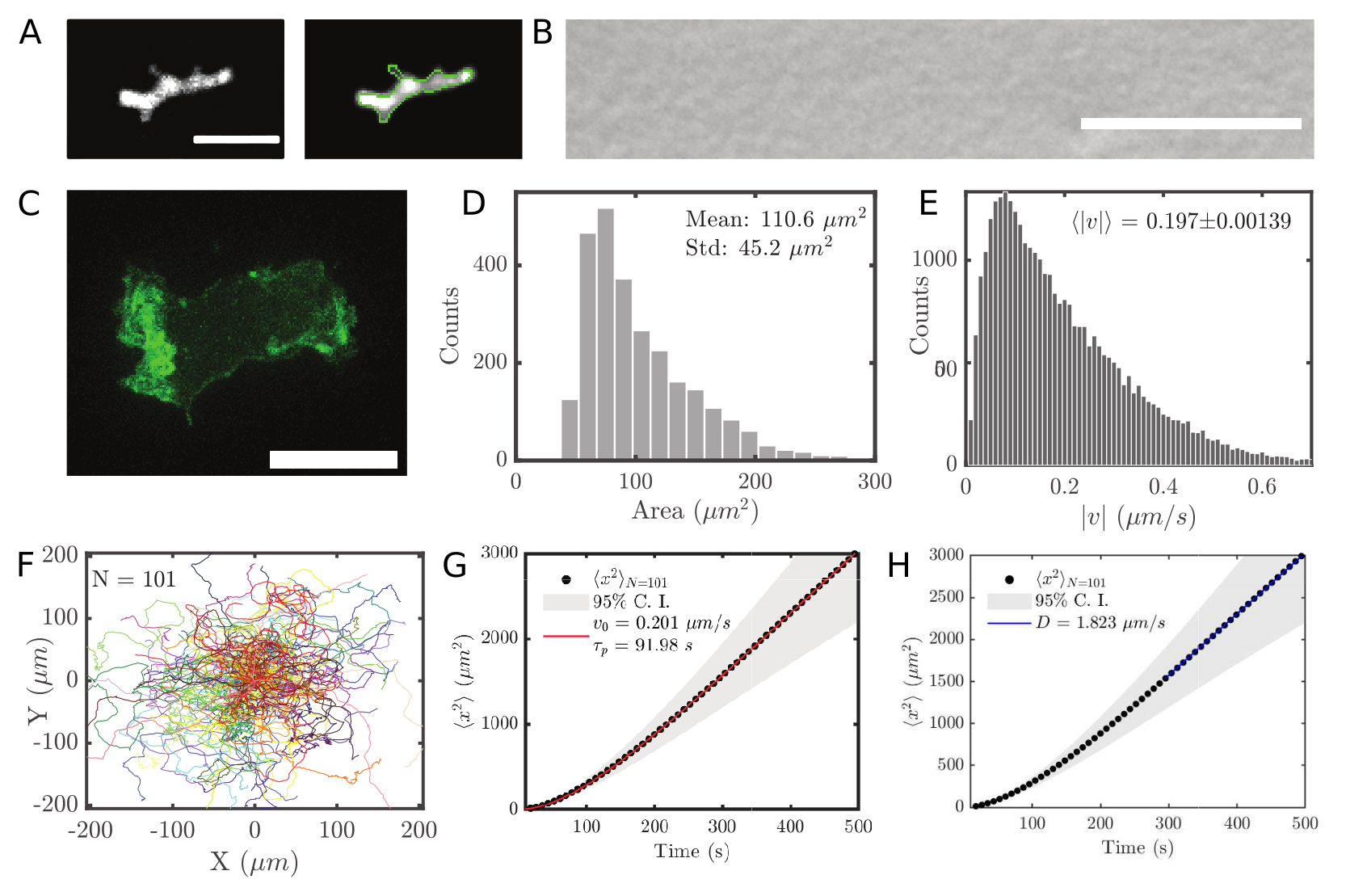}
		\caption{\textbf{Empirical basis of CPM parameter constraints: starved \textit{D. discoideums}}\\
		(A) Detection of starved \emph{D. discoideum} motion on flat PDMS was used to calibrate the CPM simulations. Time-lapse fluorescent confocal imaging of cells (left) was used to detect cell edges (green, right). Scale bar is  10\si{\micro\metre}.
		(B) SEM image of flat PDMS. Without engineered topographies, the substrate has nanometer scaled features providing traction to cells. Scale bar is 1 \si{\micro\metre}
		(C) Actin polymerization hot spots (visualized through LimE-GFP, z-projection) are highly anisotropically distributed, associated with the high persistence of starved \textit{D. discoideum}. See also \nameref{mov:actvisstarved}
		(D) Areas measured for starved cells on flat PDMS. These determined the area ($A_{\sigma}$) parameter used in the CPMs.
		(E) Instantaneous speed distribution of observed cell motility.
		(F) Trajectories obtained from live cell imaging.
        (G) The MSD (black) of starved \emph{D. discoideum} trajectories is well fit by analytical expression for persistent random particles (red). The average instantaneous speed ($\langle |v| \rangle$) and persistence time ($\tau_p$) were used to calibrate the ABP model and the CPMs.
        (H) The MSD (black) of starved \emph{D. discoideum} trajectories at larger lag-times is well fit by a line (blue). The slope of the blue line is used to derived the diffusion constant $D$, which was used to calibrate the Act-CPM and ABP-CPM.}
        \label{fig:TuningCPM}
	\end{figure}

\begin{figure}
    \includegraphics[width=\textwidth]{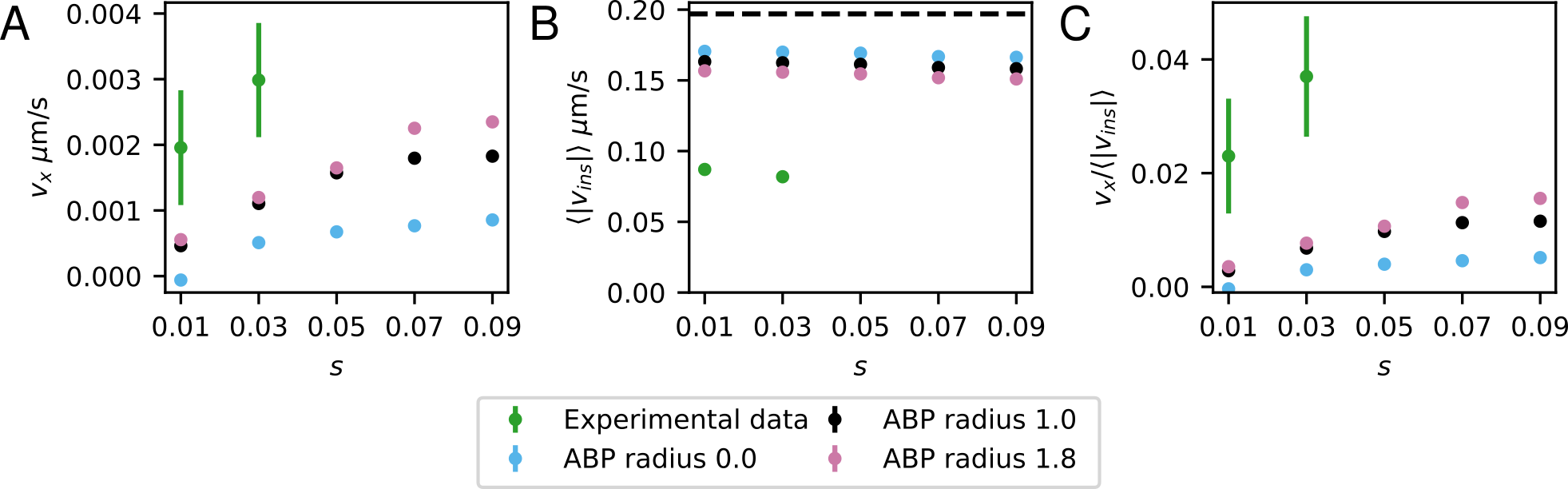}
    \caption{\textbf{Migratory drifts and instantaneous speed of ABP simulations with starved parameter set against gradient steepness $s$ for different particle radii.}
    Each data point represent the average of a 10000 simulations. Experimental data is shown for reference. Error bars indicate 97.5\% CI.
    (A) Migratory drifts $v_x$.
    (B) Instantaneous speed $\langle |v_{ins}|\rangle$.
    (C) Relative drift $v_x/\langle |v_{ins}|\rangle$.
}
    \label{fig:abp_r}
\end{figure}

\begin{figure}
    \centering
    \includegraphics[height=0.9\textheight]{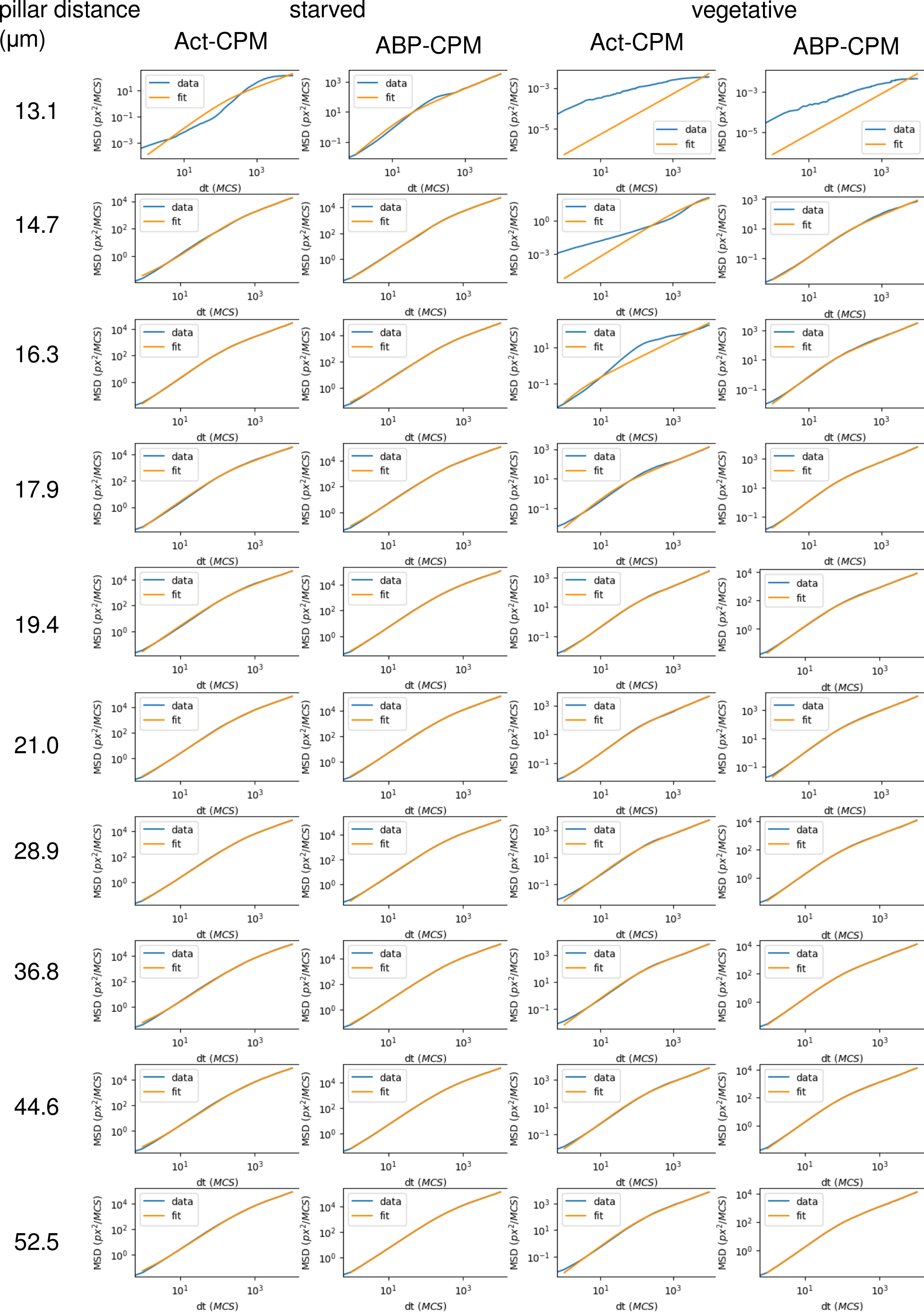}
    \caption{\textbf{MSD and fitted Eq.~\ref{eq:furthtransdiff} against time-lag $dt$ for the CPM models in regularly spaced pillar grids} Columns: Act-CPM or ABP-CPM for both starved and vegetative parameter settings. Rows: pillar distance $d$}
    \label{fig:fits}
\end{figure}

\begin{figure}
    \centering
    \includegraphics[width=\textwidth]{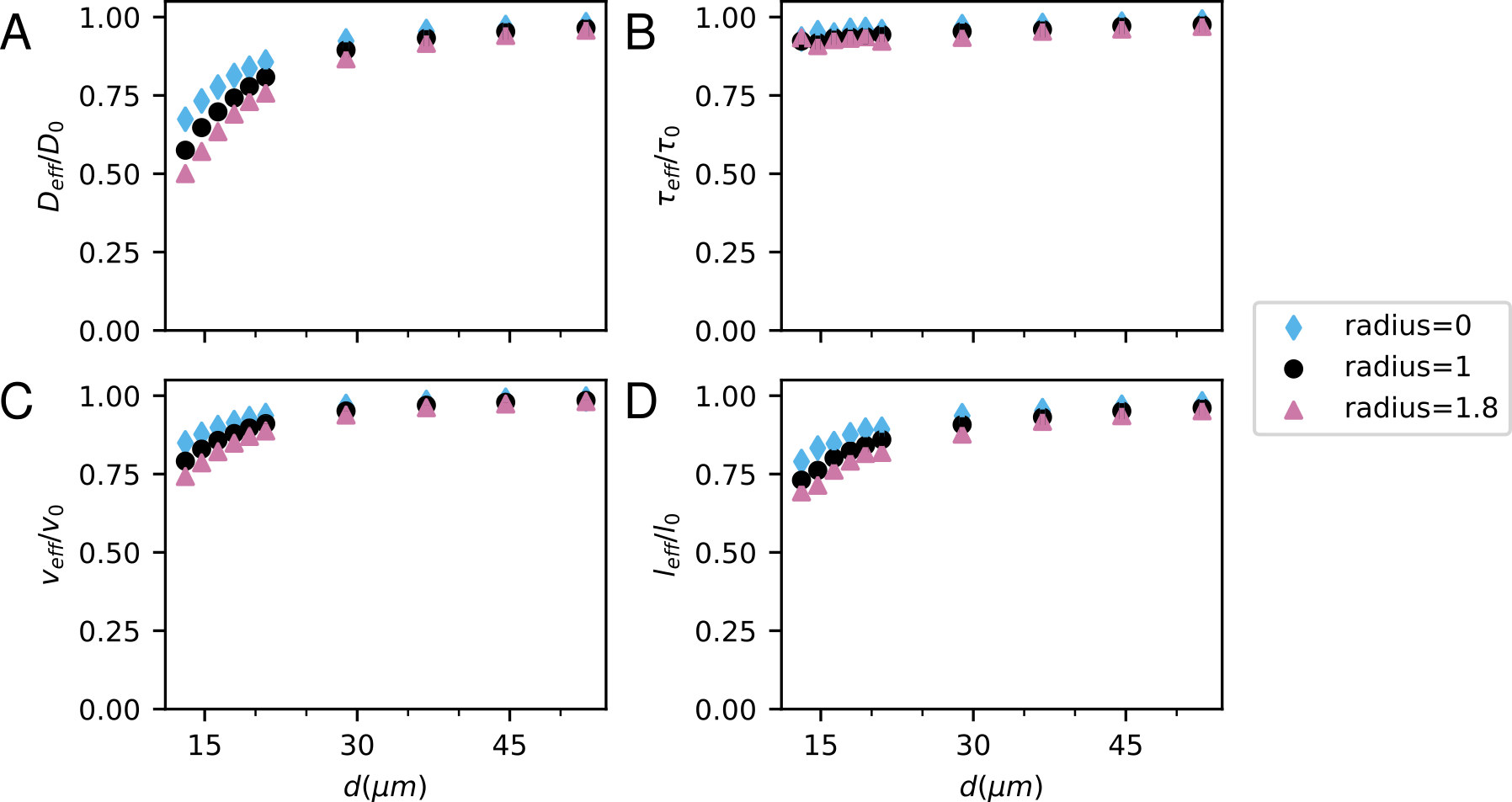}
    \caption{\textbf{Normalized effective parameters of the persistent random walk in regular pillar grids for the ABP model with the starved parameter settings for different radii of the particle}
    (A) Effective diffusion coefficient $D_{\text{eff}}$ normalized by the diffusion coefficient of starved simulations on an empty field as a function of the distance $d$ between pillar centers.
    (B) Effective persistence time $\tau_{\text{eff}}$ normalized by the persistence time of starved simulations on an empty field as a function of $d$.
    (C) Effective instantaneous speed $v_{\text{eff}}$ normalized by the instantaneous speed of starved simulations on an empty field as a function of $d$.
    (D) Effective persistence length $l_{\text{eff}}=v_{\text{eff}}\tau_{\text{eff}}$ normalized by the persistence length of starved simulations on an empty field as a function of $d$. Error bars indicate 97.5\% CI.}
    \label{fig:reg_abp}
\end{figure}

\clearpage
\appendix
\titleformat{\section}{\sffamily\bfseries\large\raggedright}{Appendix \thesection}{1em}{\MakeTextUppercase}
\section{Topotaxis of vegetative \textit{D. discoideum} cells}\label{sec:vegetative}
\renewcommand{\thefigure}{A\arabic{figure}}
\renewcommand{\thetable}{A\arabic{table}}

\setcounter{figure}{0}
\subsection*{Baseline motility of vegetative \textit{D. discoideum}}
\begin{figure*}[h]
		\centering
		\includegraphics[width=\textwidth]{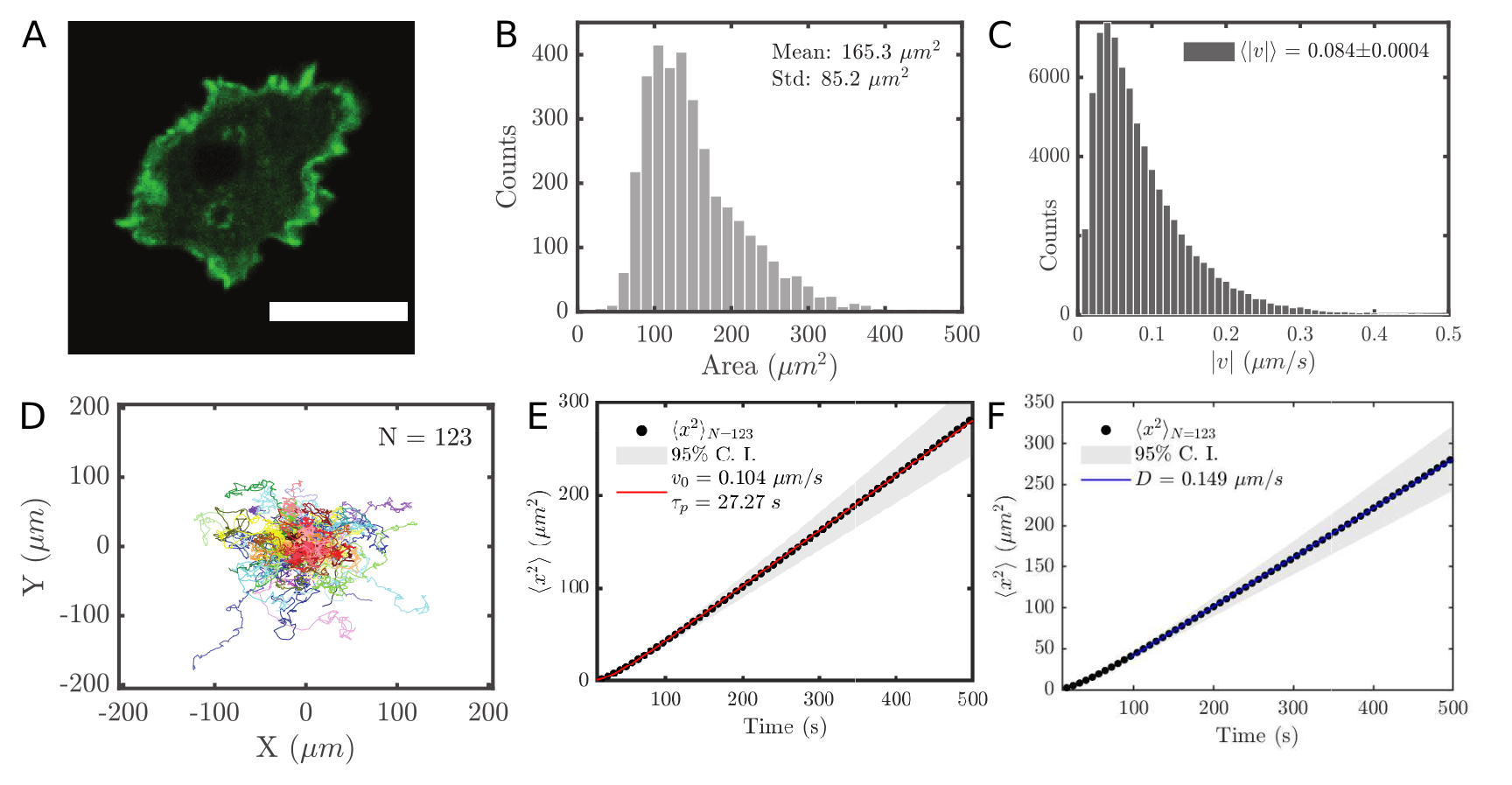}
		\caption{\textbf{Empirical basis of CPM parameter constraints: vegetative \textit{D. discoideums}}
		(A) Actin polymerization hot spots (visualized through LimE-GFP, z-projection) in the cell are more isotropically distributed in the vegetative state, giving rise to slower and less persistent motion than starved \textit{D. discoideum}.
		(B) Areas measured for vegetative cells on flat PDMS. These determined the area ($A_{\sigma}$) parameter used in the CPMs.
		(C) Instantaneous speed distribution of observed cell motility.
		(D) Trajectories obtained from live cell imaging, motion detection was performed equal to the starved state (see Fig.~\ref{fig:TuningCPM}).
        (E) The MSD (black) of starved \emph{D. discoideum} trajectories is well fit by analytical expression for persistent random particles (red, but the parabolic regime for low lag-times is significantly shorter. The average instantaneous speed ($\langle |v| \rangle$) and persistence time ($\tau_p$) were used to calibrate the ABP model and the CPMs.
        (F) The MSD (black) of starved \emph{D. discoideum} trajectories at larger lag-times is well fit by a line (blue). The slope of the blue line is used to derived the diffusion constant $D$, which was used to calibrate the Act-CPM and ABP-CPM. \label{fig:TuningCPMII}}
	\end{figure*}

In order to match \textit{in silico} cell motility to \textit{in vitro} cell motily, we determined a number of \textit{in vitro} cell properties for vegetative \textit{Dictyostelium discoideum} on a flat PDMS substrate (Fig.~\ref{fig:TuningCPMII}, \nameref{mov:actvisveg}). We measured the mean cell area (Fig.~\ref{fig:TuningCPMII}B), mean speed (Fig.~\ref{fig:TuningCPMII}C), persistence time (Fig.~\ref{fig:TuningCPMII}E) and diffusion constant (Fig.~\ref{fig:TuningCPMII}F), summarized in Table~\ref{tab:fitsII}. In general, vegetative \textit{D. discoideum} cells have a larger area and are less motile and less persistent compared to starved \textit{D. discoideum}.

To fit the parameters $\lambda_{Act},\text{Max}_{Act}$ and $\lambda_\text{persistence},\tau_p$, we followed the same procedure as for the starved cells, with the note that we increased the target area $A_\sigma$ and perimeter $P_\sigma$ to accommodate for the larger cell area of vegetative \textit{D. discoideum}, such that the resulting spatial scales between the starved and vegetative parameter sets remained similar. The motility properties and the corresponding parameter that were the best fit are shown in Tables \ref{tab:fitsII} and \ref{tab:parII}. Example simulations can be seen in \nameref{mov:simvegflat}.

\begin{table*}
\caption{Cell motility properties for vegetative \textit{D. discoideum} on a flat PDMS surface and best fits for the ABP model, Act-CPM and ABP-CPM.}
\centering
\begin{tabular}{l|llll}
                      & Experimental     & ABP model &Act-CPM        & ABP-CPM   \\ \hline
Length equivalent 1px (\si{\micro\metre})& -             & -   &  0.524   &  0.524     \\
Time equivalent 1 MCS (\si{s}) & -              & -    &  0.388       &    0.821      \\
Speed (\si{\micro m/s})      & 0.084\textpm 0.0004  & 0.079
  & -          &         -                \\
Area (\si{\micro m^2})      & 164.6\textpm 84.3    & - &  -        &   -            \\
Effective diffusion (\si{\micro m^2/s})    & 0.149\textpm 4.89$\times 10^{-4}$  & 0.097	\textpm0.002
 & 0.137\textpm0.007        &   0.121\textpm0.006       \\
Persistence time (\si{s})      & 27.41\textpm0.41   & 27.64\textpm0.55             &    25.48\textpm0.35    &  33.27\textpm1.48 \\
\end{tabular}

\label{tab:fitsII}
\end{table*}

\begin{table}
\caption{Parameter values of the best fits for the vegetative \textit{Dictyostelium} cells for the Act-CPM and ABP-CPM.}
\centering
\label{tab:parII}
\begin{tabular}{l|l|l|l}
Parameter                  & Vegetative & Parameter                    & Act-CPM \\ \hline
T                          & 20         & $\lambda_{Act}$              & 80      \\
$A_\sigma$                      & 600        & $\text{Max}_{Act}$           & 33      \\ \cline{3-4}
$\lambda_\text{area}$      & 50         & Parameter                    & ABP-CPM \\ \cline{3-4}
$P_\sigma$                  & 350        & $\lambda_\text{persistence}$ & 78      \\
$\lambda_\text{perimeter}$ & 1          & $\tau$                       & 7       \\
$J_\text{cell,medium}$     & 20         &                              &
\end{tabular}
\end{table}

\subsection*{CPM predicts small topotactic drift for vegetative cell topotaxis, but independent of persistence method}
Although no experimental data is available on topotaxis by vegetative cells in these pillar gradients, we are still interested in whether vegetative cells are able to perform topotaxis. The largest difference in motility between starved and vegetative cells is the lower speed and lower persistence time in the vegetative cells (Tables~\ref{tab:fits},\ref{tab:fitsII}), and hence, a lower persistence length of about 2.3 \si{\micro\metre}. This is smaller than the smallest gap between pillars in our pillar grids. Hence, we are unsure if collisions can reduce persistence length and thus whether topotaxis happens for this type of motility. Here, we predict whether topotaxis by vegetative cells is possible.

\begin{figure*}
    \centering
    \includegraphics[width=\textwidth]{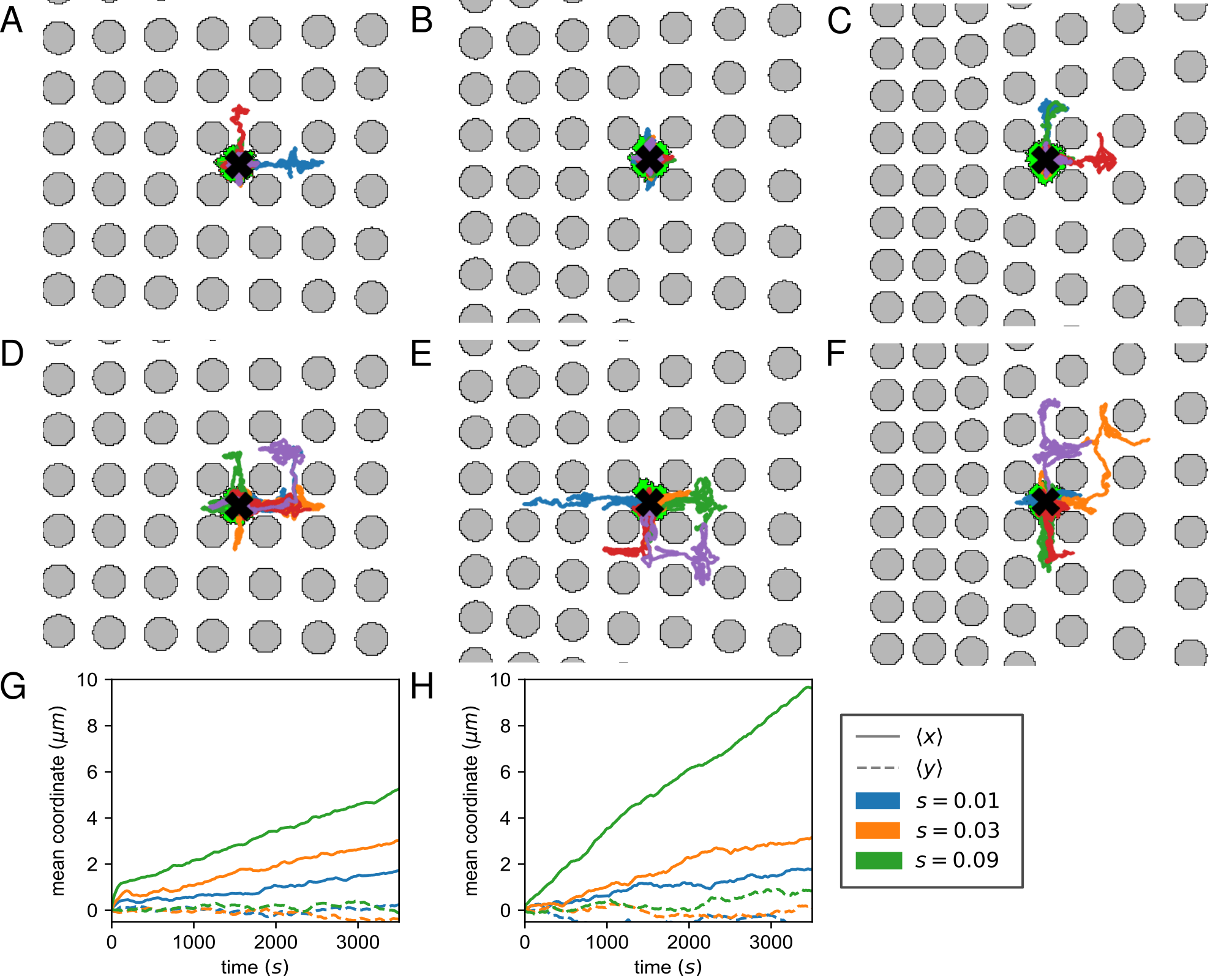}
    \caption{\textbf{Trajectories  of  vegetative  simulated  cells and means of $x$ and $y$ coordinates over time for different gradient steepness $s$.}  \emph{Top row}: Act-CPM trajectory examples:
    (A) $s= 0.01$,
    (B) $s= 0.03$,
    (C) $s= 0.09$. \emph{Middle row}: ABP-CPM model. Trajectory examples:
    (D) $s= 0.01$,
    (E) $s= 0.03$,
    (F) $s= 0.09$. For the trajectory plots: starting location is marked with a black cross and the depiction of a cell. Each trajectory has its own color.
    \emph{Bottom row} mean $x$ and $y$ coordinates of 1000 independent simulations.
    (G) Act-CPM.
    (H) ABP-CPM.}
    \label{fig:examples_vegetative_r}
\end{figure*}

The CPM cells reflect the lower motility of vegetative \textit{D. discoideum} (Fig.~\ref{fig:examples_vegetative_r}, \nameref{mov:simvegpil}) as they only move in the order of magnitude of 1-3 cell lengths on the same time scale as in Figs.~\ref{fig:r_examples},\ref{fig:J_examples}. Still, there is a visible increase in the mean $x$-coordinate but not mean $y$-coordinate (Fig.~\ref{fig:examples_vegetative_r}), so the CPMs predict that vegetative \textit{Dictyostelium} cells can perform topotaxis as well.

\begin{figure*}
    \centering
    \includegraphics[width=\textwidth]{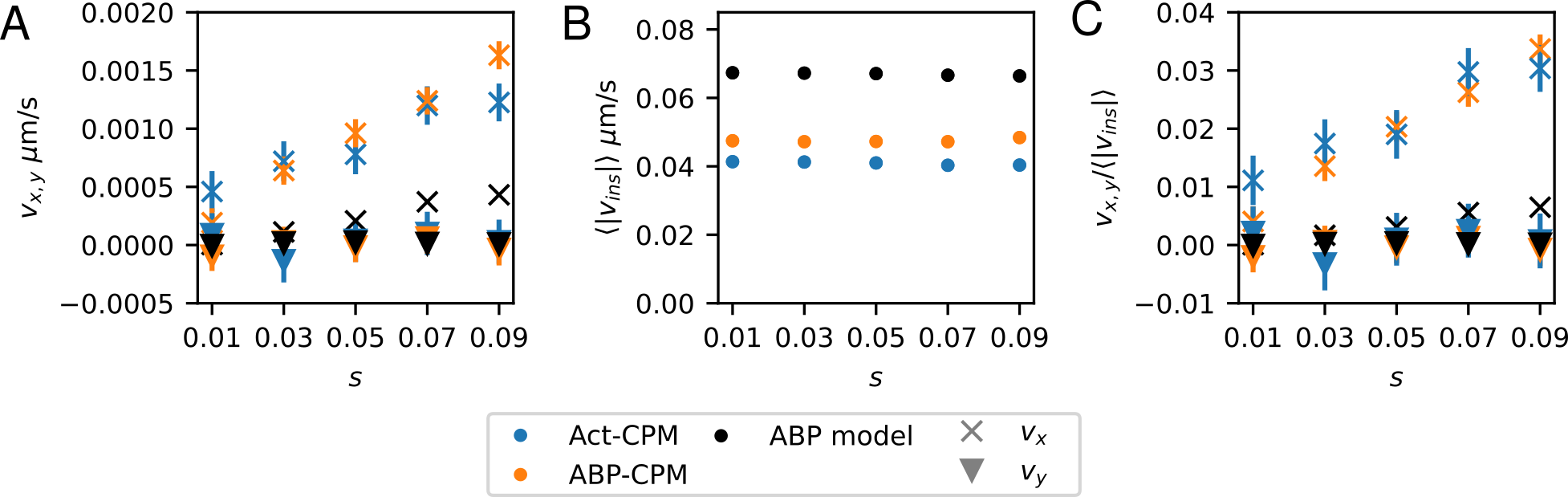}
    \caption{\textbf{Migratory drifts and instantaneous speed of vegetative parameter set against gradient steepness $s$.}
    Each data point represent the average of a 1000 simulations. Error bars indicate 97.5\% CI.
    (A) Migratory drifts $v_x$ (crosses) and $v_y$ (triangles).
    (B) Instantaneous speed $\langle |v_{ins}|\rangle$.
    (C) Relative drifts $v_x/\langle |v_{ins}|\rangle$ (crosses) and $v_y/\langle |v_{ins}|\rangle$ (triangles).
}
    \label{fig:vegetative_r}
\end{figure*}

\begin{figure*}
    \centering
    \includegraphics[width=\textwidth]{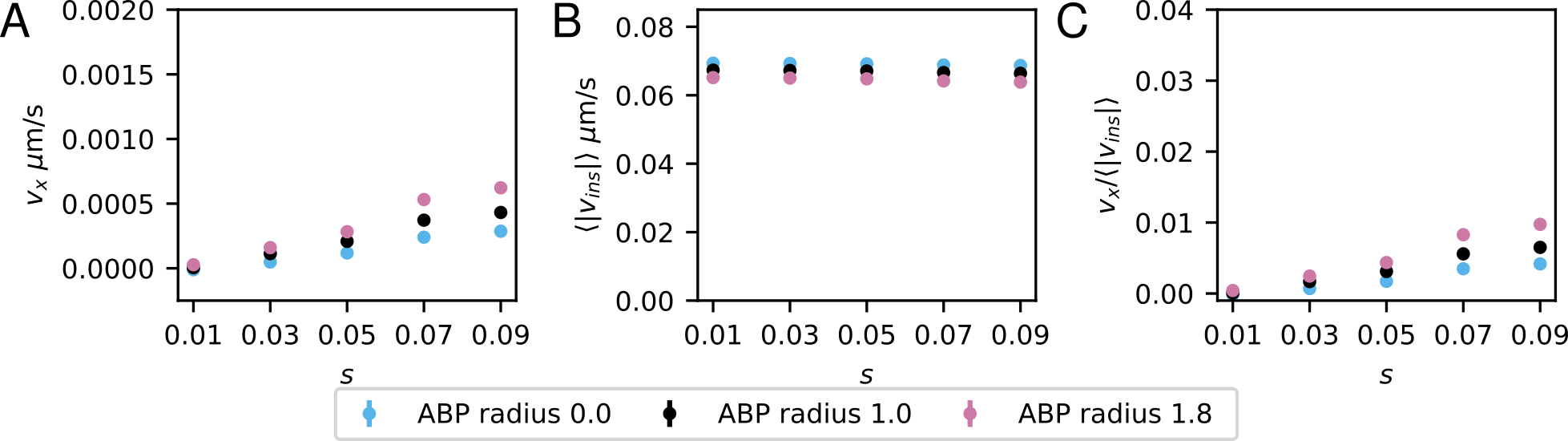}
    \caption{\textbf{Migratory drifts and instantaneous speed of ABP simulations with vegetative parameter set against gradient steepness $s$ for different particle radii.}
    Each data point represent the average of a 10000 simulations. Error bars indicate 97.5\% CI.
    (A) Migratory drift $v_x$.
    (B) Instantaneous speed $\langle |v_{ins}|\rangle$.
    (C) Relative drift $v_x/\langle |v_{ins}|\rangle$.
}
    \label{fig:abp_r_veg}
\end{figure*}

% Tests opnieuw doen
We also measured the topotactic drift and speed of the CPMs and ABP model for different values of $s$ (Figure~\ref{fig:abp_r_veg}. With the lowered displacement of vegetative cells, also $v_x$ is an order of magnitude smaller than that of starved cells (Fig.~\ref{fig:vegetative_r}A). Still, for all $s$, the topotactic drift in both CPMs exceeds the drift in the $y$-direction, (largest student's t-test $p$-values for $s=0.01$: $p=0.001$ for Act-CPM, $p=0.0002$), whereas this doesn't hold for the ABP model at $s<0.05$ (student's t-test $p=0.87$ for $s=0.01$ and  $p=0.065$ for $s=0.03$). We expect that it will be very hard to measure topotaxis in the experimental system with vegetative cells, as the starved cell CPM simulations have much smaller confidence intervals for $v_x$ than those of the experimental starved cells. Further, we again observe that the topotactic drift is proportional to $s$, and that both CPMs show more topotaxis than the ABP model. In contrast to the starved CPM simulations, the two persistence methods result in similar $v_x$ (Fig.~\ref{fig:vegetative_r}A). As CPM's $v_{ins}$ are again similar (Fig.~\ref{fig:vegetative_r}B), this results in similar relative topotactic velocity as well (Fig.~\ref{fig:vegetative_r}C). Additionally, we see a higher $v_{ins}$ for the ABP model than the CPMs, which we did not see for the starved parameter set and could be due to the increased area of the vegetative parameter set. Testing different particle radii (Figure~\ref{fig:abp_r_veg}) also indicate that increased volume results in lowered speed yet more topotaxis.
% Interestingly, the relative topotactic drift of vegetative cells lies in the range of those of starved ABP-CPM (Fig.~\ref{fig:starved_r}C).

\begin{figure*}
    \centering
    \includegraphics[width=\columnwidth]{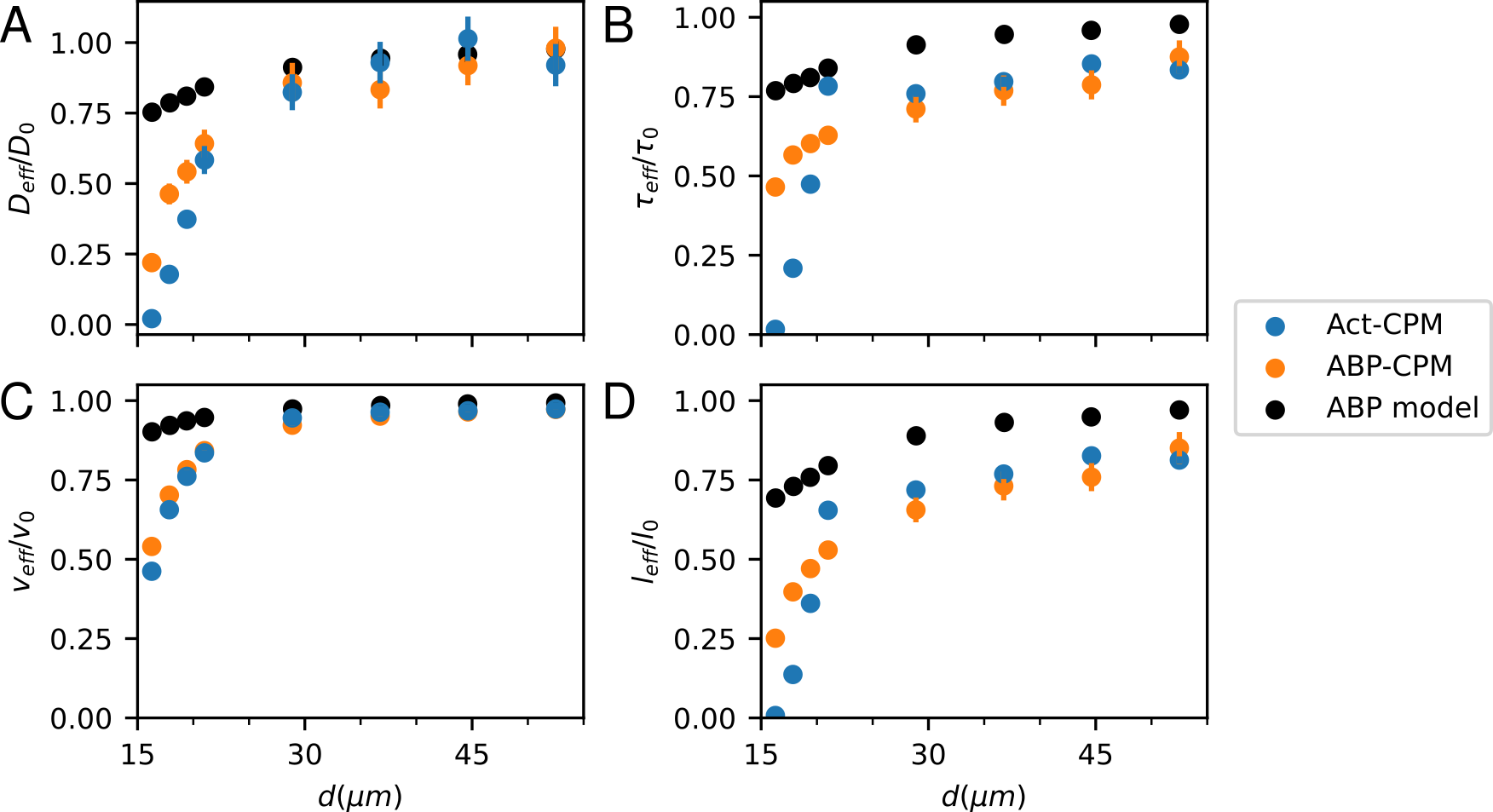}
    \caption{\textbf{Normalized effective parameters of the persistent random walk in regular pillar grids for the Act-CPM, ABP-CPM and ABP model with the vegetative parameter settings.}
    (A) Effective diffusion coefficient $D_{\text{eff}}$ normalized by the diffusion coefficient of vegetative simulations on an empty field as a function of the distance $d$ between pillar centers.
    (B) Effective persistence time $\tau_{\text{eff}}$ normalized by the persistence time of vegetative simulations on an empty field as a function of $d$.
    (C) Effective instantaneous speed $v_{\text{eff}}$ normalized by the instantaneous speed of vegetative simulations on an empty field as a function of $d$.
    (D) Effective persistence length $l_{\text{eff}}=v_{\text{eff}}\tau_{\text{eff}}$ normalized by the persistence length of vegetative simulations on an empty field as a function of $d$. Error bars indicate 97.5\% CI.}
    \label{fig:reg_veg}
\end{figure*}

We also investigated the behavior of the models with the vegetative parameters in regular grids. Again, we checked their effective diffusion $D_\text{eff}$, effective instantaneous speed $v_\text{eff}$, effective persistence time $\tau_\text{eff}$ and effective persistence length $l_\text{eff}$ as a function of distance $d$ between pillar centers in relation to these parameters on an empty field, similar to previously done for the starved cells. Again, we obtained $\tau_{\text{eff}}$ by fitting Eq.~\ref{eq:furthtransdiff} to the MSD. We observed that for the vegetative active reorientation model, Eq.~\ref{eq:furthtransdiff} is not a good descriptor of the MSD for low pillar distance $d=13.1$ \si{\micro\metre} for both CPMs up to $d=16.3$ \si{\micro\metre} for the Act-CPM (Figure~\ref{fig:fits}). Hence, we only show the results in Figure \ref{fig:reg_veg} for $d\geq16.3$ \si{\micro\metre}. The effective parameters are more similar between the two persistence methods than observed for the starved parameter set (Fig.~\ref{fig:reg}). Only on the smallest displayed pillar distances $16.3\leq d\leq19.4$\si{\micro\metre}, we see that the effective persistence time of the Act-CPM is affected more than that of the ABP-CPM. Compared to the differences between the persistence methods for the starved cells, this difference here is smaller in both relative and absolute manner. Furthermore, both CPMs show a decreased $\tau_\text{eff}$ compared to the ABP model that we did not see for the starved parameter set. As this has happens for the ABP-CPM as well, we think this decrease is due to the increased cell area. Overall, these effective parameters do not indicate a clear difference between the persistence methods for vegetative cells.

\newpage
\section*{Supporting Movies}
Movies can be found on \url{https://tbb.bio.uu.nl/leonie/dcirectdm_supp_mov/Movies.zip}
\paragraph{Movie S1}\label{mov:actvispil} Visualization of actin polymerization (starved cell) during pillar collisions in the pillar grid. The cell was imaged every 4 seconds by 10 \SI{1}{\micro\metre} Z-stacks (max-projected in movie). The pillar field was imaged after the time-lapse and used as a background.  Scale bar is \SI{10}{\micro\metre}.
    \paragraph{Movie S2}\label{mov:actvisstarved} Visualization of actin polymerization for starved \textit{D. discoideum} (Ax2 LimE GFP in Lim0) on flat PDMS substrate. The cell was imaged every 5 seconds by 10 \qty{1}{\micro\metre} Z-stacks (max-projected in movie). Scale bar is \SI{10}{\micro\metre}.

    \paragraph{Movie S3}\label{mov:simstarvedflat} Simulations of starved \textit{D. discoideum} on empty field for both Act-CPM and ABP-CPM.
    \paragraph{Movie S4}\label{mov:simstarvedpil} Simulations of starved \textit{D. discoideum} in pillar grid with $s=0.03$ for both Act-CPM and ABP-CPM.
    \paragraph{Movie S5}\label{mov:actvisveg} Visualization of actin polymerization for vegetative \textit{D. discoideum} (Ax2 LimE GFP in Lim0) on flat PDMS substrate. The cell was imaged every 5 seconds by 10 \SI{1}{\micro\metre} Z-stacks (max-projected in movie). Scale bar is \SI{10}{\micro\metre}.
    \paragraph{Movie S6}\label{mov:simvegflat} Simulations of vegetative \textit{D. discoideum} on an empty field for both Act-CPM and ABP-CPM.
    \paragraph{Movie S7}\label{mov:simvegpil} Simulations of vegetative \textit{D. discoideum} in pillar grid with $s=0.03$ for both Act-CPM and ABP-CPM.

% Uncomment if using biblatex
% \printbibliography
\end{document}